\documentclass[prl,amsmath,amssymb,twocolumn,superscriptaddress]{revtex4}
\usepackage{amsmath,amssymb,graphicx}
\usepackage{mathrsfs}
\usepackage{dcolumn}
\usepackage{bm}

\begin{document}

\title{Unconventional Relaxation of Hydrodynamic Modes in Anharmonic Chains}

\author{Daxing Xiong}
\email{xmuxdx@163.com}
\email{phyxiongdx@fzu.edu.cn}
\affiliation{School of Science, Jimei University, Xiamen 361021, Fujian,
China}
\affiliation{Department of Physics, Fuzhou University, Fuzhou
350108, Fujian, China}

\begin{abstract}
Nonlinear fluctuating hydrodynamics (NFHD) is a powerful framework for understanding transport, but checking its validity with molecular dynamics is still challenging. Here, we overcome this challenge by developing an effective scheme for detecting hydrodynamic modes that takes into account the role of pressure fluctuations. We show that the predictions given by NFHD on the relaxation processes of hydrodynamic modes are valid only when the pressure of the system is zero and the pressure fluctuations are weak. For nonvanishing pressure, two other regimes arise as the hydrodynamic modes can respond to small and large pressure fluctuations and relax in two additionally distinct manners. In contrast to the previous finding of two classes, our results suggest that there are at least three universality classes of transport in anharmonic chains.
\end{abstract}

\maketitle
\emph{Introduction}.---Transport by normal diffusion can be decomposed into several decoupled hydrodynamic modes which relax exponentially toward the equilibrium state. This is the main prediction given by linear fluctuating hydrodynamics (FHD)~\cite{FHD}, a stochastic fluid dynamics approach, that describes fluctuations by usual hydrodynamic equations supplemented with random noises~\cite{Noteadd}. More concretely, it pointed out that a fluctuation can induce three hydrodynamic modes: two propagating sound modes and one standing heat mode; all three are decoupled and with diffusive broadening. During the past decades this linear FHD has become a powerful tool for modeling microscopic fluctuations in three-dimensional systems of a manner consistent with statistical mechanics and nonequilibrium thermodynamics. Nevertheless, in one spatial dimension, it fails as the transport is usually superdiffusive~\cite{Longtail}. Hence, understanding anomalous transport from a nonlinear version of FHD~\cite{FHD-1,Narayan2002,Beijeren2012,Spohn2014} is currently a central fundamental issue.

So far several efforts combining FHD with renormalization group analysis~\cite{Narayan2002}, mode-coupling theory~\cite{Beijeren2012} and nonlinear Langevin equation~\cite{Spohn2014} have been made. This is essentially a second-order correction to FHD, named nonlinear fluctuating hydrodynamics (NFHD). Actually, it was first devoted to understanding anomalous thermal conduction in anharmonic chains, a typical paradigm of intensive interest~\cite{Book1,LepriReport,DharReport}. Here, the anomaly means a breakdown of the Fourier law $J=- \kappa \nabla T$, which describes heat current $J$ proportional to temperature gradient $\nabla T$ of a constant thermal conductivity $\kappa$, replaced by a power-law system size ($L$) dependent $\kappa \sim L^{\gamma}$~$(0<\gamma<1)$. In NFHD, this anomaly is related to heat mode with superdiffusive broadening. In this respect, two generic universality classes of $\gamma =1/2$ and $1/3$ for transport, classified by vanishing and nonvanishing system's pressures, respectively, have been predicted.

However, in spite of these efforts, it is still inclusive whether and under what conditions the central predictions of the NFHD hold. The NFHD claimed that, to capture the superdiffusive broadening, a second-order correction suffices and all other descriptions of its linear version, like the decoupling hypothesis of hydrodynamic modes, are still valid. This was, indeed, partially verified in the Fermi-Pasta-Ulam (FPU) chains by a suggested scheme~\cite{Spohn2014,NumTest}. Nevertheless, this scheme requires the averaged pressure as an input at the starting point, which is not as explicitly available as for anharmonic chains~\cite{Spohn2014}. This causes the scheme unfavorable to study chains under strong pressure fluctuations~\cite{NFHD-no-3}. On the other hand, recent progress on generalized hydrodynamics suggested that even for noninteracting, integrable systems, a higher-order hydrodynamics is required~\cite{Generalized-1,Generalized-2}. All of these then question the validity of NFHD on general grounds~\cite{Livi2020}. It is thus of fundamental importance to propose a more favorable scheme. This would help make new findings, just like the seminal discoveries of long-time tails made by Alder and Wainwright~\cite{Longtail}.

In this Letter we develop a variant, effective scheme to capture hydrodynamic modes in anharmonic chains on general grounds. An obvious advancement here is to include the role of pressure fluctuations. Using this scheme we show that the central predictions of NFHD work only in a class of systems of zero pressure and weak pressure fluctuations. As the pressure becomes nonzero, two other regimes with unexpected relaxations of hydrodynamic modes emerge. These three regimes suggest that there are at least three universality classes for transport, contrary to the previous two classes. The result sheds new light on how to further improve the NFHD theory.

\emph{Anharmonic chains}.---The Hamiltonian of a general anharmonic chain is
\begin{equation} \label{Hamiltonian}
H= \sum_{\mu=1}^{L} p_{\mu}^2/2 + V(r_{\mu+1}-r_\mu),
\end{equation}
where $p_\mu$ is the $\mu$th (totally $L$ particles and all with unit mass following periodic boundary conditions) particle's momentum, $r_{\mu}$ is its displacement from equilibrium position, and $V(\xi)$ is the interparticle potential. We mainly consider a typical cubic-plus-quartic chain with $V(\xi)= {\alpha \xi^3}/3+\xi^4/4$ $(\alpha \geq 0)$~\cite{Xiong2018,Lee-Dadswell2008}. This model highlights the effects of nonlinearity and thus helps us explore common properties of an anharmonic chain on general grounds (see Supplemental Material~\cite{SM}). The particular case of $\alpha=0$ corresponds to a purely quartic chain whose $V(\xi)$ (averaged pressure) is symmetric (zero), where the universality class of $\gamma=1/2$ has been conjectured~\cite{Beijeren2012,Spohn2014}. A nonvanishing $\alpha$ then adds potential's asymmetry (equivalently, the averaged pressure will mathematically become nonzero) and makes $\gamma$ belong to another universality class of $\gamma=1/3$. Such a variation of $\gamma$ has not yet been fully verified from the relaxation of hydrodynamic modes.

\begin{figure}[!t]
\vskip-0.6cm
\includegraphics[width=8.8cm]{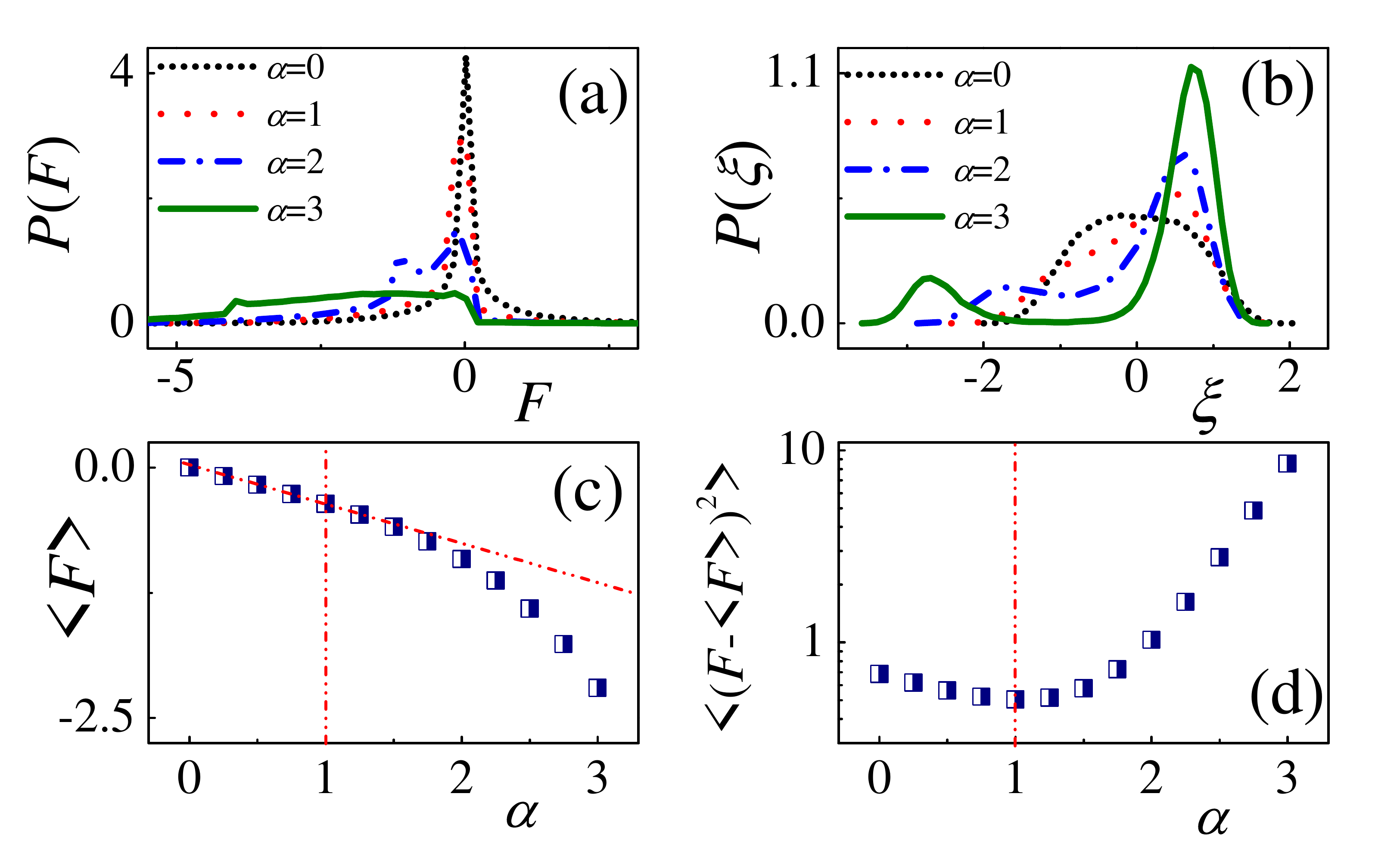}
\vskip-0.7cm \caption{The cubic-plus-quartic chains (the same below
for Figs.~2$-$4): Distribution of (a) pressure and (b) displacement
($\xi=r_{\mu+1}-r_\mu$) under equilibrium ($T=0.5$) for different
$\alpha$. Accordingly, (c) and (d) give the averaged pressure $\langle F
\rangle$ and its fluctuation $\langle (F- \langle F \rangle )^2
\rangle$ vs $\alpha$.} \label{fig:1} \vskip-0.4cm
\end{figure}

\emph{Pressure and pressure fluctuations}.---The full role of pressure ($F$) can be seen from its distribution $P(F)$ and fluctuations $\langle (F- \langle F \rangle )^2 \rangle$ in the equilibrium state ($T=0.5$). As shown in Fig.~\ref{fig:1}, only for $\alpha=0$ or a relatively small $\alpha$ $(\leq 1)$, $P(F)$ indicates a narrow peak [see Fig.~\ref{fig:1}(a)], which leads the displacement distribution $P(\xi)$ that denotes the potential to mainly concentrate on a single well [see Fig.~\ref{fig:1}(b)]. As $\alpha$ increases, both $P(F)$ and $P(\xi)$ appear in a wide range and seem to be around two peaks. Further examinations of the averaged pressure $\langle F \rangle$ and $\langle (F- \langle F \rangle )^2 \rangle$ reveal that $\alpha_{\mathrm{cr}} \simeq 1$ seems to be a turning point [see Figs.~\ref{fig:1}(c,d)], i.e., below (above) which the $\alpha$-dependence of $\langle F \rangle$ is linear (nonlinear) and the value of $\langle (F- \langle F \rangle )^2 \rangle$ is small (large). Such a turning point, even exhibiting locally, has also been explored by studying systems' dynamic structure factor~\cite{Xiong2018}, indicating its nonlocal property. Due to this variation of structure, it is necessary to check the validity of NFHD by carefully considering the role of pressure fluctuations.

\emph{Numerical Scheme}.---Unfortunately, there has not yet been an effective scheme to derive hydrodynamic modes on general grounds. As mentioned, the previous scheme requires the pressure as an input~\cite{NumTest}, but the pressure itself is not explicitly available, which causes us unable to consider the effects of pressure fluctuations. To overcome this shortcoming, we suggest that one can return to the original FHD formalism~\cite{FHD-1,Beijeren2012} and seek out an effective method.

Specializing the FHD to one dimension, the relevant hydrodynamic modes are defined by small wave number $k$ and given as linear combinations of the Fourier components of the deviations from the equilibrium values of the microscopic densities of particle $\hat{n}(k,t)=\sum_{\mu=1}^L \exp(-i k r_\mu)-\delta_{k,0} \langle \hat{n} \rangle$; momentum $\hat{p}(k,t)=\sum_{\mu=1}^L p_\mu(t) \exp(-i k r_\mu)$, and energy $\hat{e}(k,t)=\sum_{\mu=1}^L e_\mu(t) \exp(-i k r_\mu)-\delta_{k,0} \langle \hat{e} \rangle$. Here $\delta$ represents the Kronecker delta function, $\langle \cdot \rangle$ denotes the ensemble average, $\langle \hat{n} \rangle$ and $\langle \hat{e} \rangle$ are the equilibrium values of the Fourier components ($k=0$) for densities of particle ($1$) and energy [$e_\mu(t)=p_{\mu}^2/2 + V(r_{\mu+1}-r_\mu)/2+V(r_{\mu}-r_{\mu-1})/2$]. In fact, such three quantities are conserved fields of the system and it is assumed that all slow variables relevant to the long-time behavior of hydrodynamics and related time correlations are the long-wavelength Fourier components of densities of these conserved fields plus their product~\cite{Beijeren2012}. Under this assumption, the hydrodynamic modes are one heat mode:
\begin{equation}
\hat{a}_{\sigma_0}(k,t)=\left(\frac{1}{\langle n \rangle k_B T^2
C_p}\right)^{1/2} [\hat{e}(k,t)-h \hat{n}(k,t)]
\end{equation}
and two sound modes:
\begin{equation} \label{SoundF}
\hat{a}_{\sigma_{\pm}} (k,t)=\left(\frac{1}{2 \langle n \rangle
T}\right)^{1/2} [c_0^{-1} \hat{F}(k,t)+\sigma_{\pm} \hat{p}(k,t)],
\end{equation}
with $\sigma_0$ and $\sigma_{\pm}=\pm 1$, respectively, labeling the heat and sound; $k_B$ the Boltzmann constant; $\hat{F}(k,t)=\sum_{\mu=1}^L F_\mu(t) \exp(-i k r_\mu)-\delta_{k,0}
\langle \hat{F} \rangle$ the pressure fluctuation with $F_\mu(t)=-\partial V(r_{\mu+1}-r_\mu)/\partial r_\mu$ (in anharmonic chains the pressure is equal to the force exerted on the particle); $C_p$ the specific heat under constant averaged pressure $\langle F \rangle$; $h=(\langle e \rangle + \langle F \rangle)/\langle n \rangle$ the equilibrium enthalpy per particle, and $c_0$ the sound speed. As seen, both $\hat{a}_{\sigma_0}$ and $\hat{a}_{\sigma_{\pm}}$ already explicitly involve the information of pressure, which is distinct from the previous scheme~\cite{Spohn2014}. More importantly, the pressure fluctuations are key to sound modes since from Eq.~\eqref{SoundF} these modes are combinations of pressure and momentum densities. Now since the Fourier transform is a linear transformation, in real space one obtains:
\begin{equation} \label{heat}
a_{\sigma_0}(l,t)=\left(\frac{1}{\langle n \rangle k_B T^2 C_p}\right)^{1/2} [\Delta e_l(t)-h \Delta n_l(t)];
\end{equation}
\begin{equation} \label{sound}
a_{\sigma_{\pm}} (l,t)=\left(\frac{1}{2 \langle n \rangle
T}\right)^{1/2} [c_0^{-1} \Delta F_l(t)+\sigma_{\pm} \Delta p_l(t)]
\end{equation}
with $\Delta \mathscr{X}= \mathscr{X}_l(t)-\langle \mathscr{X} \rangle$ the fluctuations of the relevant quantities.

We here replace the lattice index $\mu$ by a coarse-grained space number $l$. To do this one can decompose the chain into several equal slabs (each contains $s=L/L_s$ particles with $L_s$ the total size of the slabs). In this sense $\mathscr{X}_l(t)=\sum_{\mu \in l}\mathscr{X}_\mu(t)$ is a coarse-grained description of $\mathscr{X}$ via a slab. This facilitates the measurement of $n_l(t)$'s fluctuations; otherwise $\Delta n_\mu(t)$ will vanish. One also notices that both prefactors in Eqs.~\eqref{heat} and~\eqref{sound} are constants for a given temperature. Therefore, even these prefactors always appear, to measure hydrodynamic modes correlations, one only needs to concern with  $Q_l(t)=e_l(t)-h n_l(t)$ and $S^{\pm}_l(t)=c_0^{-1}F_l(t)+\sigma_{\pm} p_l(t)$. Indeed, $Q_l(t)$ is the heat density evolved from a process. Its equilibrium time-correlation function $\rho_{Q}(m,t)=\frac{\langle \Delta Q_{l+m}(t) \Delta Q_{l}(0) \rangle}{\langle \Delta Q_{l}(0) \Delta Q_{l}(0) \rangle}$ (the translational invariance of the chain suggests that the correlation only depends on the relative distance $m$) has been employed to study heat mode in many literatures~\cite{Chen2013,Xiong2018}. In what follows our main argument is that $S^{\pm}_l(t)$ can be utilized as the densities relevant to sound and the sound modes correlations can be defined by
\begin{equation} \label{Scorrelation}
\rho_{S^{+(-)}}(m,t)=\frac{\langle \Delta S^{+(-)}_{l+m}(t) \Delta
S^{+(-)}_{l}(0) \rangle}{\langle \Delta S^{+(-)}_{l}(0) \Delta
S^{+(-)}_{l}(0) \rangle}.
\end{equation}

\begin{figure}[!t]
\vskip-0.5cm
\includegraphics[width=8.8cm]{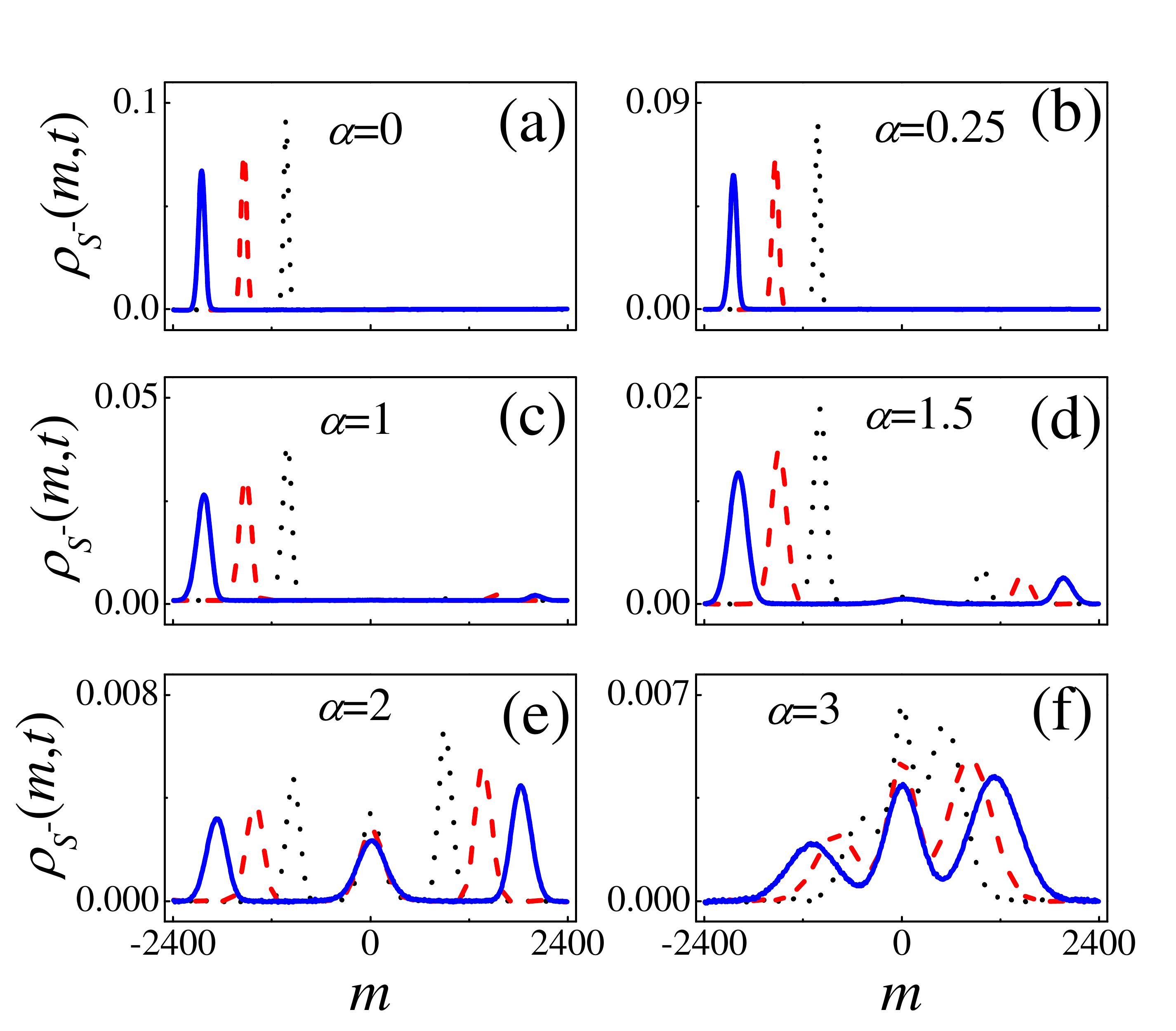}
\vskip-0.7cm \caption{Sound mode correlation $\rho_{S^{-}}(m,t)$
for three long times $t=1000$ (dotted), $t=1500$ (dashed), and
$t=2000$ (solid), for different $\alpha$.} \label{fig:2} \vskip-0.4cm
\end{figure}
\begin{figure}[!t]
\vskip-0.5cm
\includegraphics[width=8.8cm]{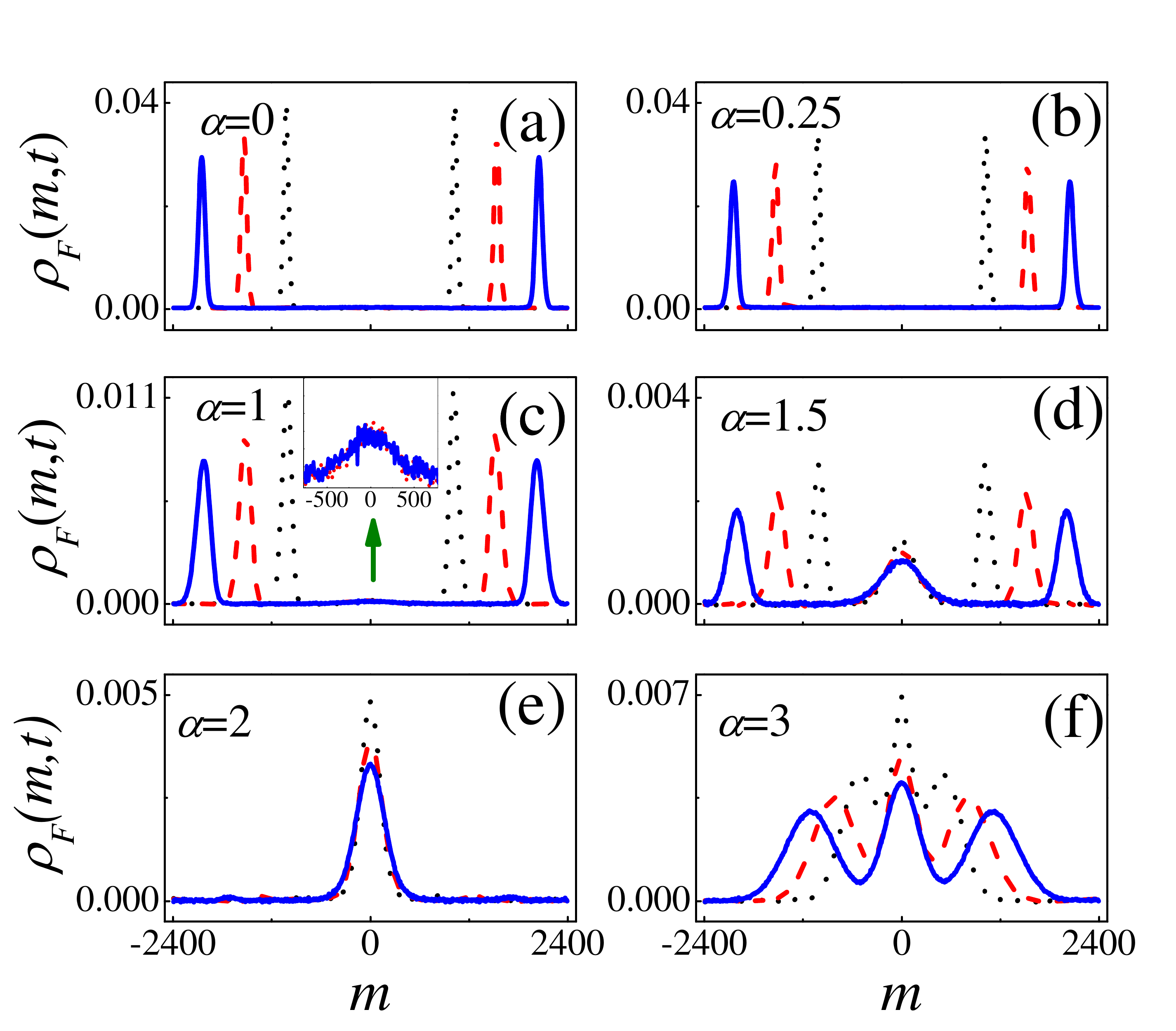}
\vskip-0.7cm \caption{The propagation of pressure fluctuations
$\rho_F(m,t)$ for three long times $t=1000$ (dotted), $t=1500$
(dashed), and $t=2000$ (solid). The inset in (c) is a zoom for the
central peaks.} \label{fig:3} \vskip-0.4cm
\end{figure}

\begin{table}[!hbp]
\begin{centering}
\vspace{-.1cm}
\begin{tabular}{c|c c c c c c}
 \hline
 \hline
  $\alpha$  & $0$ & $0.25$  & $1$ & $1.5$ & $2$ & $3$ \\ \hline
  $c_0$  & $1.022$ & $1.019$ & $1.004$ & $0.985$ & $0.918$ & $0.612$ \\ \hline
  \hline
\end{tabular}
\caption{\label{T1} The measured $c_0$ for different $\alpha$.} 
\end{centering}
\end{table}
\emph{Sound speed}.---Now in Eq.~\eqref{Scorrelation} only the sound speed $c_0$ is unknown. Generally, for a system with a given $V(\xi)$, one can obtain the thermodynamics from the pressure ensemble where $c_0$ is contained. In this respect, an analytical formula with an explicit set of inputs $V(\xi)$ and $\langle F \rangle$ has been previously suggested~\cite{Spohn2014}. However, since $\langle F \rangle$ is not explicitly available here, in our scheme we suggest that it is better to obtain $c_0$ from direct molecular dynamics. This is realized by measurement of momentum correlation $\rho_p(m,t)=\frac{\langle p_{l+m}(t) p_{l}(0) \rangle}{\langle p_{l}(0) p_{l}(0) \rangle}$~\cite{Nianbei2010}, for which two separate peaks moving ballistically with $c_0$ are always displayed~\cite{Xiong2016}. Some typical $c_0$ are listed in Table~\ref{T1}.

\emph{Sound modes}.---Figure~\ref{fig:2} depicts one branch of sound modes $\rho_{S^{-}}(m,t)$. We do not plot another branch $\rho_{S^{+}}(m,t)$ as both ones bear the mirror symmetry (see~\cite{SM}, where the detailed numerical procedures as well as the $3 \times 3$ correlator matrices of hydrodynamic modes are presented). As shown, no obvious distinctions are detected for $\alpha < 1$ [Figs.~\ref{fig:2}(a,b)], where all $\rho_{S^{-}}(m,t)$ behave similarly as predicted by NFHD, i.e., on one side there is a ballistically moving peak and this peak is broadening as time increases. This implies that the NFHD might be almost valid in this range. However, unconventional pictures emerge for a large $\alpha$, where two other peaks from both the center and the opposite sides appear [Figs.~\ref{fig:2}(d)]. These are not just small perturbations to the origin sound peak. In fact, a further increasing $\alpha$ can lead such additional peaks dominated [Figs.~\ref{fig:2}(e)]. Finally, all the three peaks are overlapped for a quite large $\alpha$ [Figs.~\ref{fig:2}(f)] indicating strong couplings. This detailed transition indicates that the conventional relaxation pictures of NFHD for sound modes might be not available for a large $\alpha$ and $\alpha_{\mathrm{cr}} \simeq 1$ seems to be a borderline [see Figs.~\ref{fig:2}(c)].

\emph{Role of pressure fluctuations}.---Indeed, $\alpha_{\mathrm{cr}} \simeq 1$ has already been shown in~Fig.~\ref{fig:1}. To further reveal the role of pressure fluctuations, one can study  $\rho_{F}(m,t)=\frac{\langle \Delta F_{l+m}(t) \Delta F_{l}(0) \rangle}{\langle \Delta F_{l}(0) \Delta F_{l}(0) \rangle}$, a correlation function that directly captures the propagation of the local pressure fluctuations along the system. As depicted in Fig.~\ref{fig:3}, $\rho_{F}(m,t)$ behaves quite similarly as sound modes: For $\alpha < 1$, two peaks appear in the location of sound modes, while for $\alpha \geq 1$, a center peak emerges [see Fig.~\ref{fig:3}(d), but Fig.~\ref{fig:3}(c) seems to be a borderline]. This peak quickly becomes dominated [see Fig.~\ref{fig:3}(e)]. Nevertheless, such a picture can not be persisted for ever, a larger $\alpha$ will lower down the peak and make all three peaks overlapped. All these are consistent with the results shown in~Fig.~\ref{fig:2}. They also suggest that $\alpha_{\mathrm{cr}} \simeq 1$ generally classifies two distinct universality classes for transport of system with nonvanishing pressure ($\alpha \neq 0$) and $\rho_{F}(m,t)$ can be used as an early indicator.

\emph{Decoupling hypothesis}.---Besides giving the sound modes, our scheme also helps test the decoupling hypothesis that was usually assumed in NFHD. This is achieved by measuring the cross-correlation
\begin{equation}
C_{S^{-}Q} (m,t)=\langle \Delta S^{-}_{l}(0) \Delta Q_{l+m}(t)
\rangle
\end{equation}
between heat and one branch of sound modes $S^{-}$. This is already enough to obtain the full coupling information as we have verified that $C_{Q S^{\pm}} (m,t)= \langle \Delta Q_{l}(0) \Delta S^{\pm}_{l+m}(t)  \rangle \equiv C_{S^{\pm}Q} (m,t) =\langle \Delta S^{\pm}_{l}(0) \Delta Q_{l+m}(t)  \rangle$ and there is a mirror symmetry between $C_{Q S^{-}} (m,t)$ and $C_{Q S^{+}} (m,t)$. For more details, one can refer to~\cite{SM}.
\begin{figure}[!t]
\vskip-0.5cm
\includegraphics[width=8.8cm]{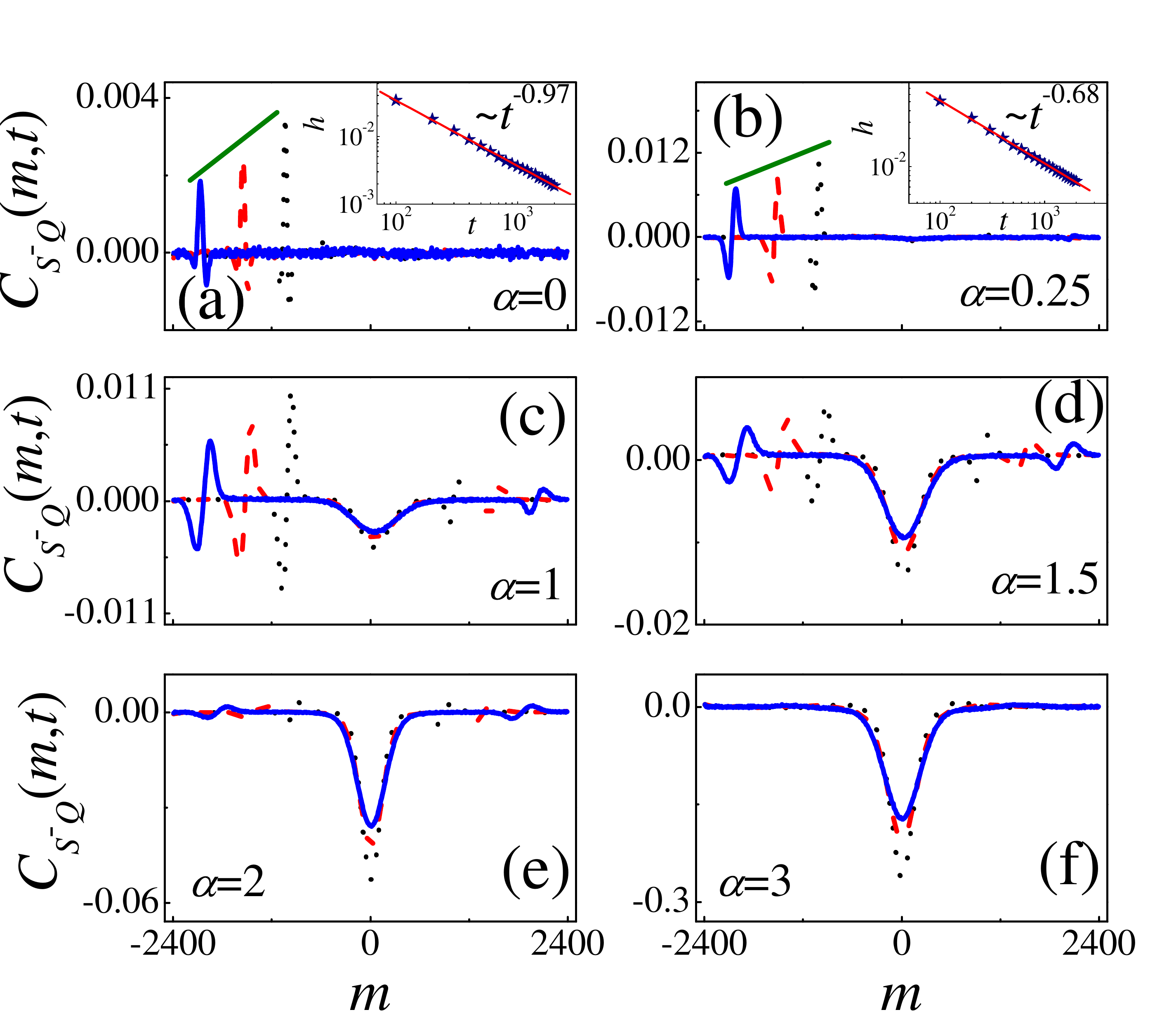}
\vskip-0.7cm \caption{The cross-correlation $C_{S^{-}Q}(m,t)$ between heat and one sound mode for three long times $t=1000$ (dotted), $t=1500$ (dashed), and $t=2000$ (solid). The insets of (a,b) depict the peaks of $C_{S^{-}Q}(m,t)$ decaying with $t$.}
\label{fig:4} \vskip-0.4cm
\end{figure}

Figure~\ref{fig:4} depicts $C_{S^{-}Q} (m,t)$. The results basically support the observed sound modes and the relevant two universality classes for $\alpha \neq 0$: For a relatively large $\alpha$, all the three places of $\rho_{S^{-}}(m,t)$ exhibit cross-correlations [see Fig.~\ref{fig:4}(d,e)]. This is the case even for the borderline $\alpha_{\mathrm{cr}} \simeq 1$ [see Fig.~\ref{fig:4}(c)]. For a quite large $\alpha$, the cross-correlations will be concentrated on the center [see Fig.~\ref{fig:4}(f)]. In this sense, the decoupling hypothesis is surely violated for $\alpha \geq 1$.
\begin{figure}
\begin{centering}
\vspace{-.5cm} \includegraphics[width=8.8cm]{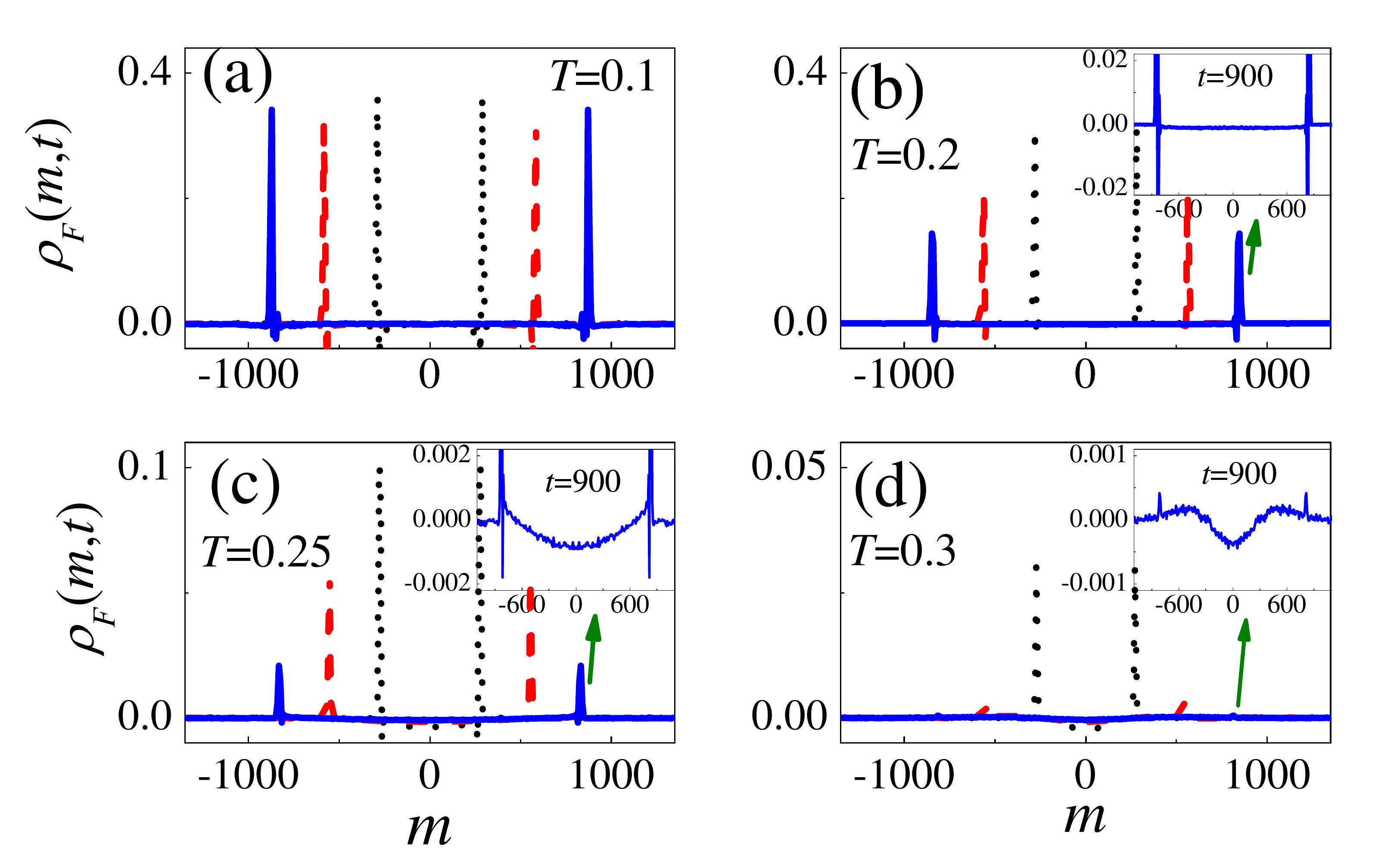} \vspace{-1.1cm}
\caption{\label{fig:5} The rotor chain: $\rho_F(m,t)$ for three
times $t=300$ (dotted), $t=600$ (dashed), and $t=900$ (solid). The
insets in (b-d) are a zoom for the result of $t=900$.}
\vspace{-.6cm}
\end{centering}
\end{figure}

Turning back to small $\alpha$, one however gains new knowledge [see Figs.~\ref{fig:4}(a,b)]. New distinctions between $\alpha=0$ (vanishing pressure) and $0<\alpha<1$ (nonvanishing pressure) emerge: For a given $t$, $C_{S^{-}Q} (m,t)$ of $\alpha=0$ looks like a single peak, but the peak for $\alpha =0.25$ is distorted. Remarkably, the single peak shown for $\alpha=0$ is different from the estimation from the previous scheme~\cite{Spohn2014} where no cross-correlations were detected~\cite{SM}. But, it is indeed consistent with the measurement of heat mode from our scheme whose the cross-correlations are slowly decaying~\cite{Xiong2018}. Another information is that the peak of $\alpha=0$ ($\alpha=0.25$) decays as $t^{-0.97}$ ($t^{-0.68}$). Obviously, here $\alpha=0$ suggests one more universality class different from $0<\alpha<1$, since its exponent is close to $-1$. Combining the similar observation in the FPU-$\beta$ chain (see~\cite{SM}), one thus infers a universal scaling $t^{-1}$ for this latter class. This fact, if compared with the diffusive broadening of sound modes (the sound peaks decay as $t^{-\frac{1}{2}}$~\cite{Spohn2014}), suggests that, the decoupling of hydrodynamic modes basically holds in a long time~\cite{Notenew}. Therefore, the NFHD is almost valid in this class. However, a longer-time tail of $C_{S^{-}Q} (m,t)$ for $\alpha=0.25$ supports that even a small deviation to this class can induce long-time couplings, thus breaking the decoupling hypothesis.

\emph{Pressure fluctuations in rotor chain}.---So far we have realized that to understand transport, besides the pressure, its fluctuations are also significant. To demonstrate the generality of these fluctuations, we lastly study $\rho_F(m,t)$ in a special rotor chain [$V(\xi)=1-\cos(\xi)$]~\cite{Rotor-1,Rotor-2,Rotor-3,Rotor-4} whose thermal conduction is conjectured to follow the Fourier law at high temperatures. In this case, sound modes are absent, leading to a diffusive heat mode. As seen in Fig.~\ref{fig:5}, the result of $\rho_F(m,t)$ shows a relevant temperature-dependent crossover from appearing to vanishing, which does demonstrate a detailed absent process of sound modes. This result thus helps answer the long-standing question why a temperature-dependent crossover from abnormal to normal transport can take place in the rotor chain~\cite{Xiong2020}.

\emph{Conclusion and discussion}.---In summary, we have proposed an effective scheme to explore hydrodynamic modes in anharmonic chains on general grounds. We have shown that in response to pressure and pressure fluctuations, unconventional relaxations with unexpected mode couplings, classified by three distinct regimes, occur. This generally results in three universality classes for thermal transport (see~\cite{SM}). The finding would be ubiquitous in general anharmonic chains with three conserved fields, as studying the FPU-$\beta$ and FPU-$\alpha$$\beta$ systems leads to similar observations~\cite{SM}. Besides, extending the scheme to study other systems like the one having more degrees of freedom~\cite{NFH3D}, or with an external pressure~\cite{NFHD-no-3} will be of interest. In this respect, the pressure fluctuations might play different roles as that was already shown in the rotor chain and new discoveries might be made. Moreover, our results can provide useful information to further improve the NFHD. For instance, the current NFHD always suggested the validity of the decoupling hypothesis~\cite{Spohn2014}. However, the strong couplings revealed here imply that necessary corrections to NFHD are required~\cite{Livi2020}.

\begin{acknowledgments}
This work was supported by NNSF (Grant No. 11575046) of China, NSF
(Grant No. 2017J06002) of Fujian Province of China. DX acknowledges useful
comments from Prof. Henk van Beijeren and the support
for attending the program - Thermalization, Many body localization
and Hydrodynamics (Code: ICTS/hydrodynamics2019/11) held in
International Centre for Theoretical Sciences (ICTS). During this
program, some valuable discussions with Prof. Herbert Spohn and
Prof. Abhishek Dhar are enjoyed. DX also appreciates Jiao Wang, Yong Zhang and Weicheng Fu for fruitful discussions in revising the manuscript.
\end{acknowledgments}


\pagebreak
\clearpage
\begin{center}
\textbf{\large Supplementary Material for
`Unconventional Relaxation of Hydrodynamic Modes in Anharmonic Chains'}
\end{center}
\setcounter{equation}{0} \setcounter{figure}{0}
\setcounter{table}{0} \setcounter{page}{1} \makeatletter
\renewcommand{\theequation}{S\arabic{equation}}
\renewcommand{\thefigure}{S\arabic{figure}}
\renewcommand{\bibnumfmt}[1]{[S#1]}
\renewcommand{\citenumfont}[1]{S#1}

\section{I. Simulation procedure}
The measurement of $\rho_p(m,t)$ as well as the hydrodynamic modes is performed as follows: We first prepare a canonical equilibrium state by employing the Langevin thermal baths for a system of size $L=7201$ to ensure an initial fluctuation located at the center to spread out at a time about $t=2000$. This can be achieved by evolving the system for a long enough time ($>10^7$ time units) from properly assigned initial random states with the baths, during which the Runge-Kutta algorithm of seventh to eighth order with a time step $0.05$ is applied. After the equilibrium states are already approached, we then remove the heat baths, evolve the system in isolation, and calculate the correlations. We use a total ensemble of size $8 \times 10^9$ for the average. In the coarse-grained description, we set $s=8$.
\begin{figure*}
\begin{centering}
\includegraphics[width=14cm]{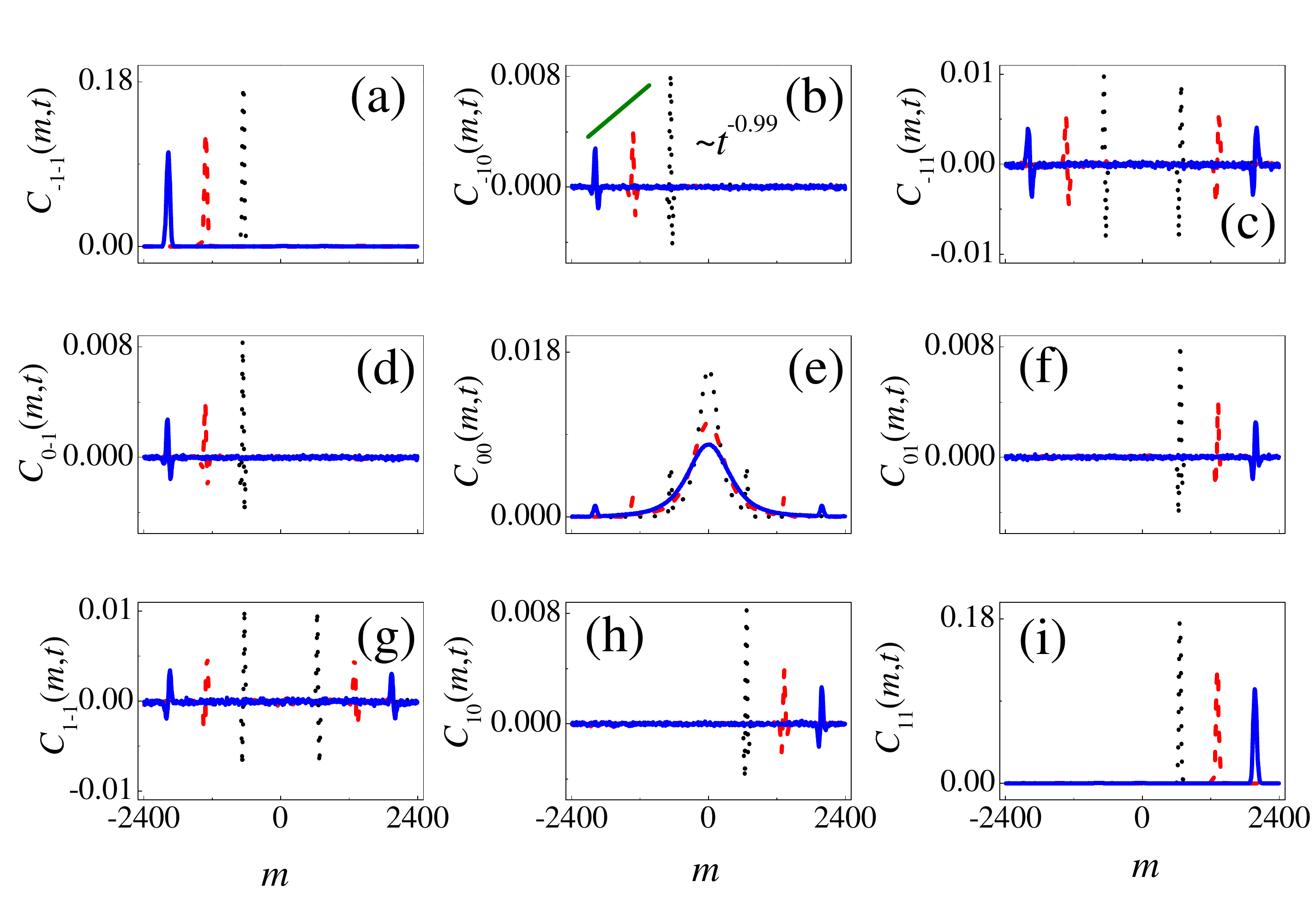} \vspace{-0.6cm}
\caption{\label{SFig1} The FPU-$\beta$ chain ($\beta=1$): the $3 \times 3$ correlator matrix of hydrodynamic modes for three long times: $t=500$ (dotted), $t=1000$ (dashed), and $t=1500$ (solid), at $T=0.5$. Here and below, for the sake of convenience, we use $\pm 1$ representing the two sound modes and $0$ the heat mode.} \vspace{-0.6cm}
\end{centering}
\end{figure*}
\section{II. FPU-$\beta$ chain}
To show that the cubic-plus-quartic chain can capture common properties of an anharmonic chain on general grounds, in this and next sections we give results  for the FPU-$\beta$ and FPU-$\alpha$$\beta$ chains. Figure~\ref{SFig1} depicts the $3 \times 3$ correlator matrix of hydrodynamic modes for the FPU-$\beta$ chain ($\beta=1$, $T=0.5$). Including the linear interaction leads to a larger sound speed, hence we only plot the results with a time up to $t=1500$. As shown, both the sound modes and their couplings to heat display similar manners as those shown in the purely quartic chain. In particular, the couplings [the peaks of $C_{-10}(m,t)$, $C_{0-1}(m,t)$, $C_{01} (m,t)$, and $C_{10} (m,t)$] decay with time as $\sim t^{-0.99}$. This, combined with the results shown in Fig. 4, seem to suggest a universal scaling law $t^{-1}$ for the class of an even potential (zero averaged pressure) and weak pressure fluctuations.
\begin{figure*}
\begin{centering}
\includegraphics[width=14cm]{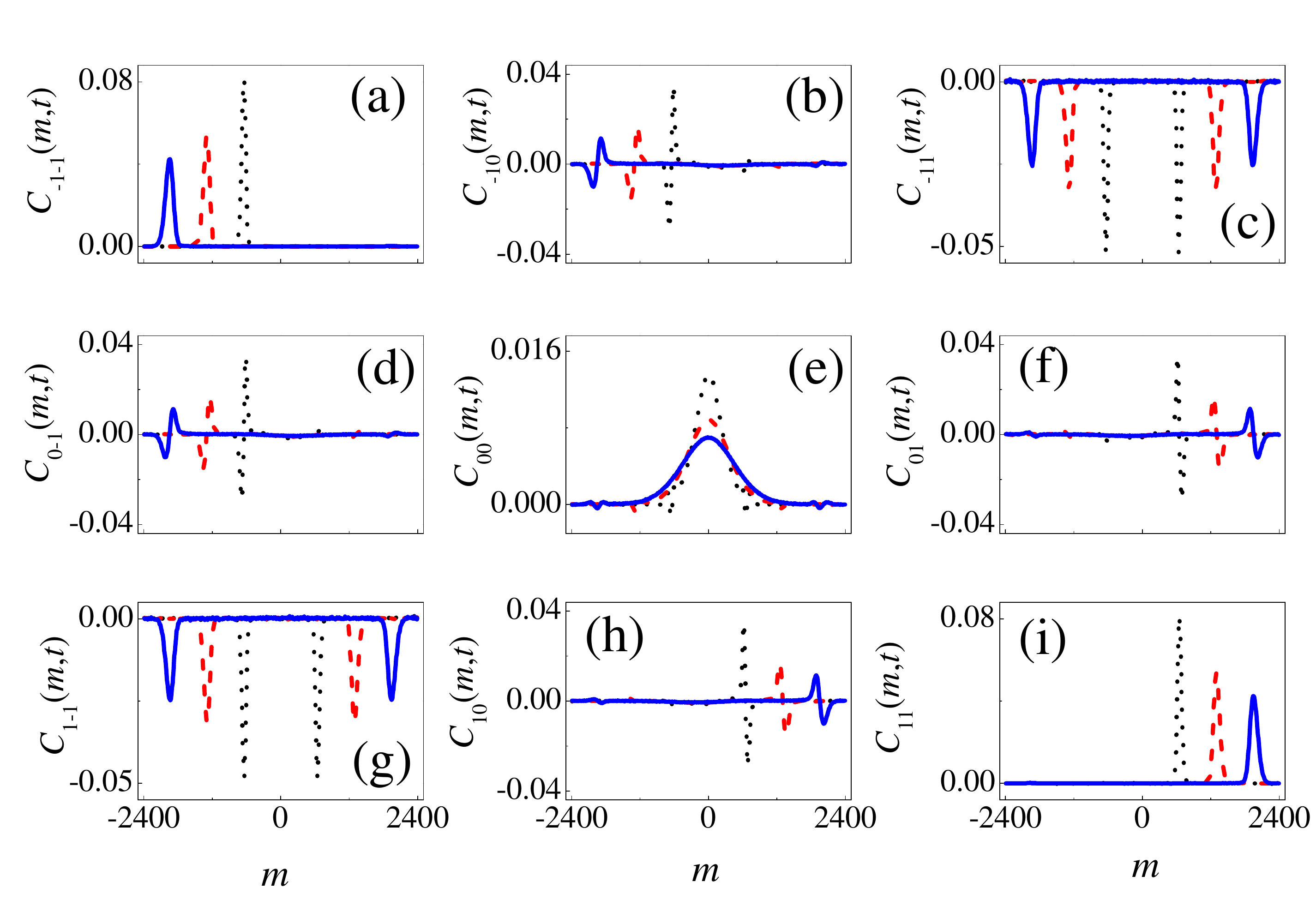} \vspace{-0.6cm}
\caption{\label{SFig2} The FPU-$\alpha$$\beta$ chain ($\alpha=1$, $\beta=1$): the $3 \times 3$ correlator matrix of hydrodynamic modes for three long times: $t=500$ (dotted), $t=1000$ (dashed), and $t=1500$ (solid), at $T=0.5$. } \vspace{-0.6cm}
\end{centering}
\end{figure*}
\begin{figure*}
\begin{centering}
\includegraphics[width=14cm]{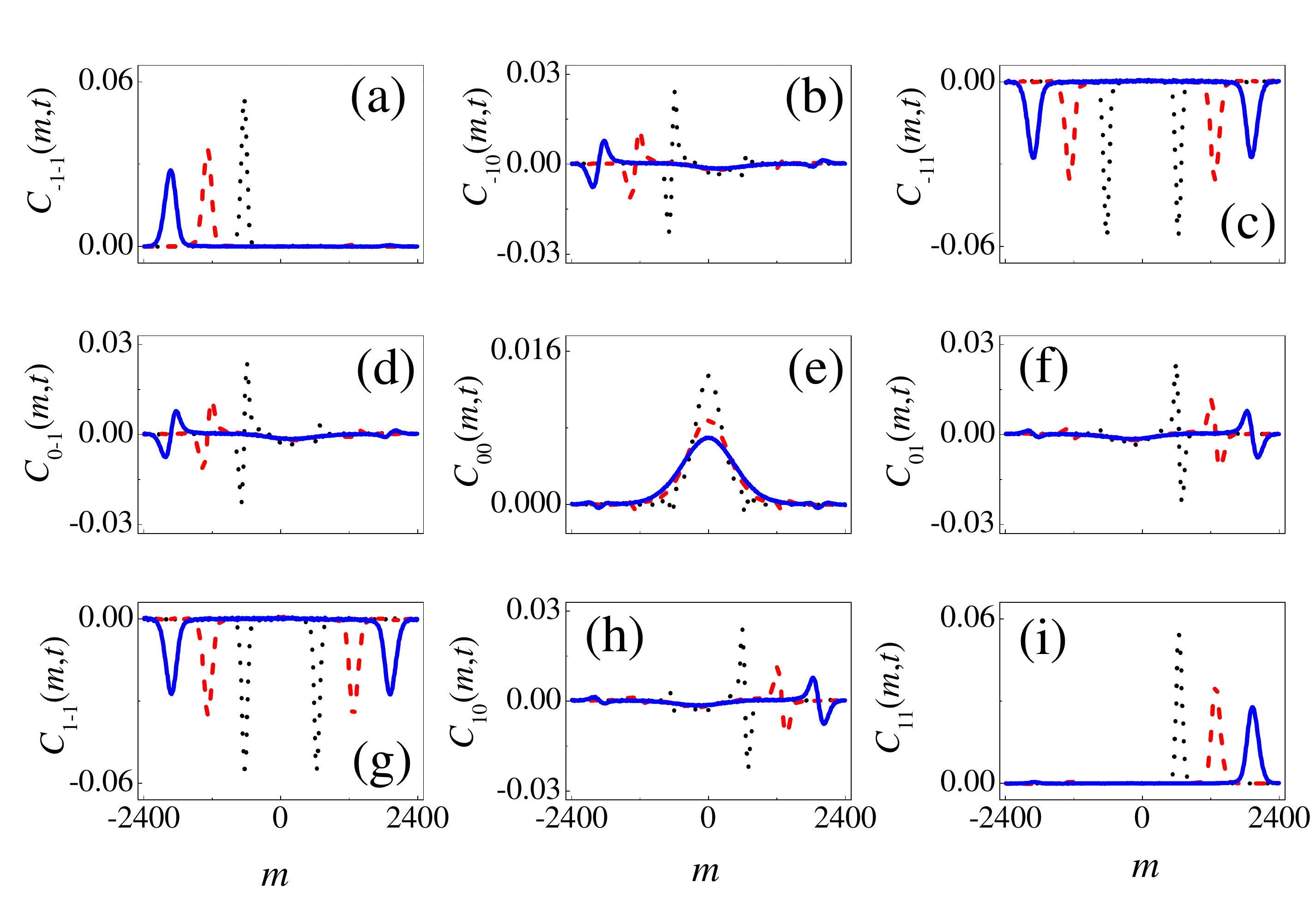} \vspace{-0.6cm}
\caption{\label{SFig3} The FPU-$\alpha$$\beta$ chain ($\alpha=1.5$, $\beta=1$): the $3 \times 3$ correlator matrix of hydrodynamic modes for three long times: $t=500$ (dotted), $t=1000$ (dashed), and $t=1500$ (solid), at $T=0.5$. } \vspace{-0.6cm}
\end{centering}
\end{figure*}
\begin{figure*}
\begin{centering}
\includegraphics[width=14cm]{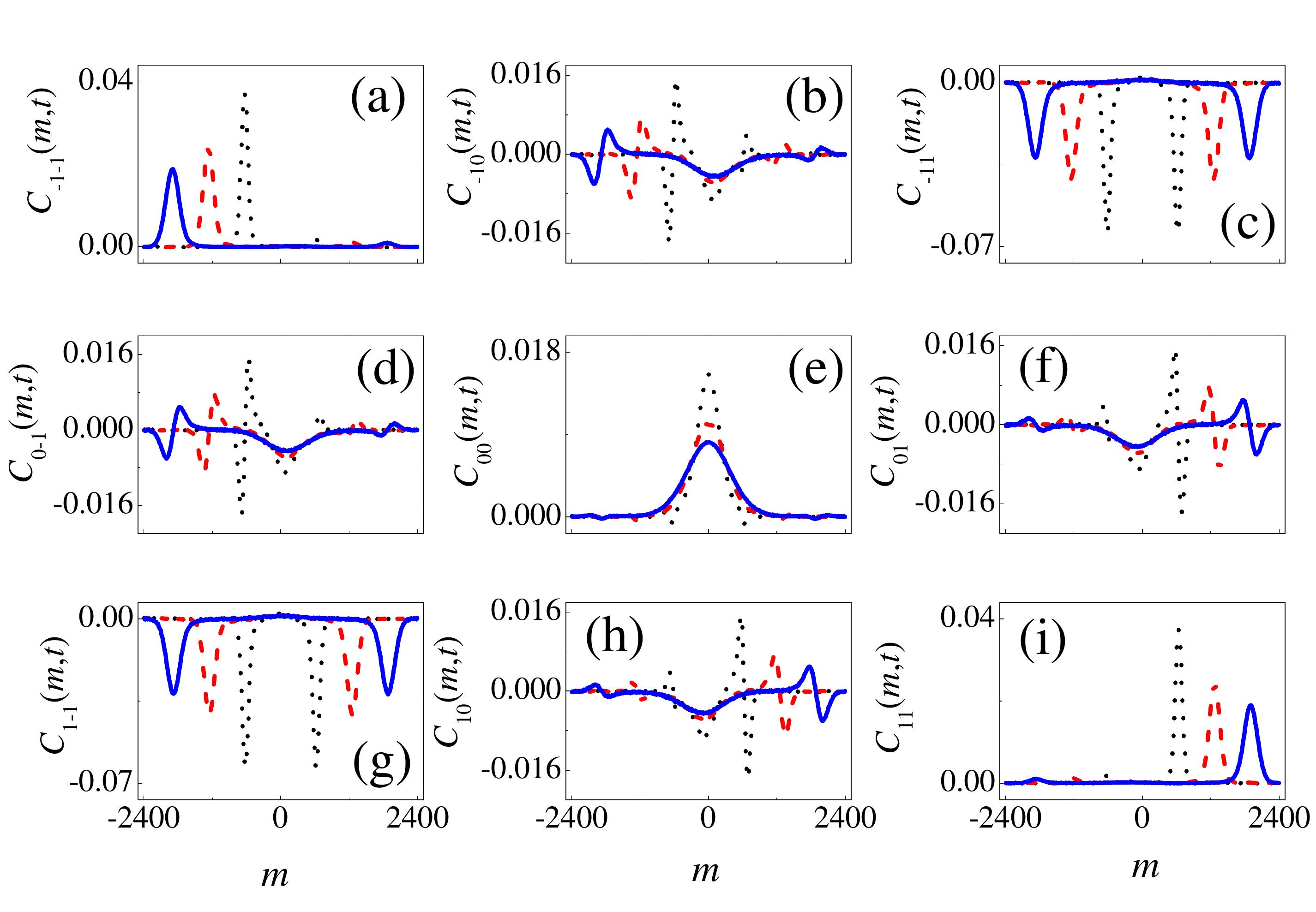} \vspace{-0.6cm}
\caption{\label{SFig4} The FPU-$\alpha$$\beta$ chain ($\alpha=2$, $\beta=1$): the $3 \times 3$ correlator matrix of hydrodynamic modes for three long times: $t=500$ (dotted), $t=1000$ (dashed), and $t=1500$ (solid), at $T=0.5$. } \vspace{-0.6cm}
\end{centering}
\end{figure*}
\section{III. FPU-$\alpha$$\beta$ chains}
Figures~\ref{SFig2},~\ref{SFig3}, and~\ref{SFig4} further depict the $3 \times 3$ correlator matrix of hydrodynamic modes for the FPU-$\alpha$$\beta$ chain with $\beta=1$ and $\alpha=1$, $1.5$, and $2$, respectively. Basically, as $\alpha$ increases they show similar manners as those displayed in the cubic-plus-quartic chains. However, including the linear interaction seems to weaken the effects of pressure fluctuations and that is why previously this unconventional relaxation has not yet been explored. Hence, in the main text it is reasonable to employ the cubic-plus-quartic chains to highlight the effects of nonlinearity.
\begin{figure*}
\begin{centering}
\includegraphics[width=14cm]{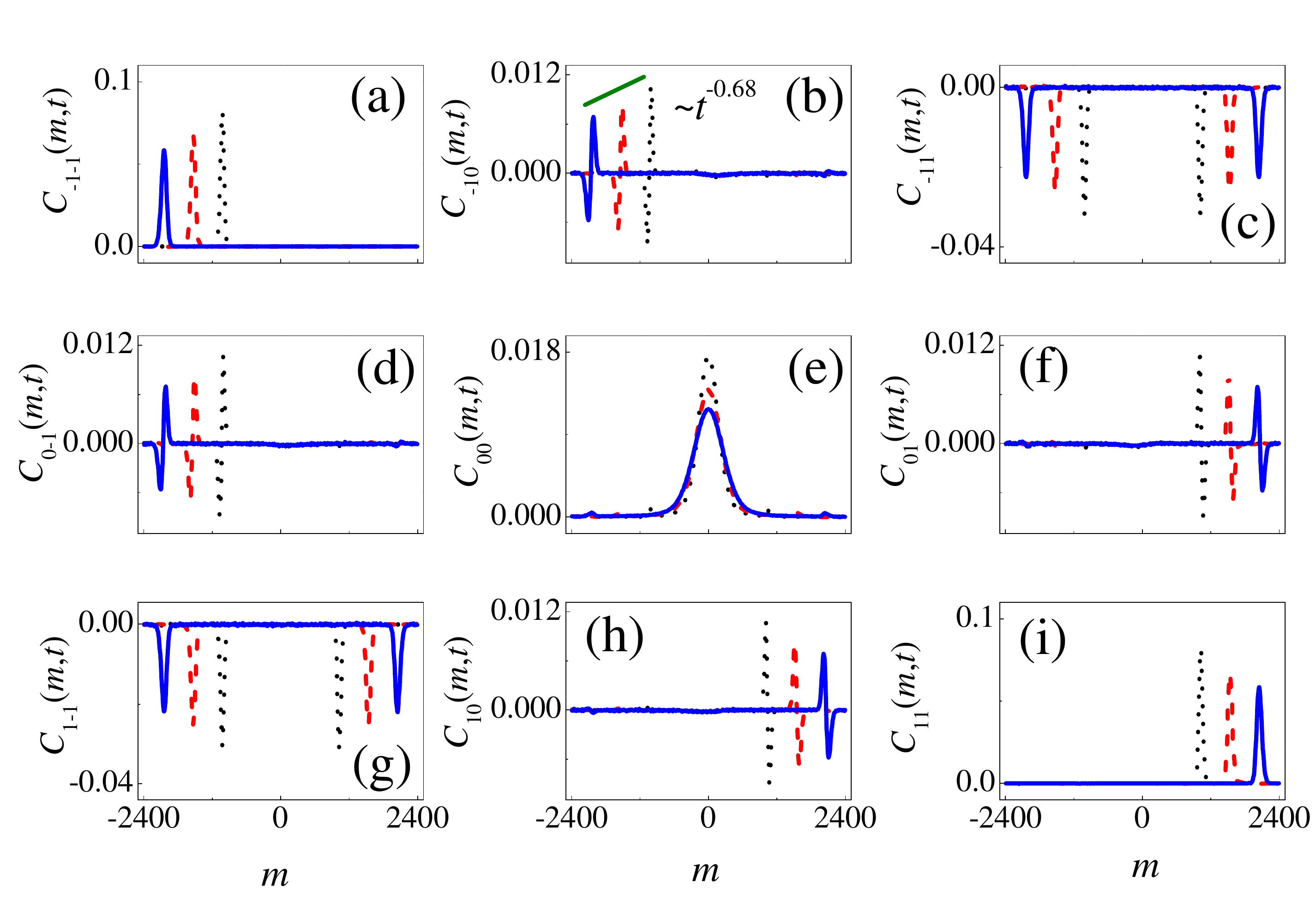} \vspace{-0.6cm}
\caption{\label{SFig5} The cubic-plus-quartic chain ($\alpha=0.25$): the $3 \times 3$ correlator matrix of hydrodynamic modes for three long times: $t=1000$ (dotted), $t=1500$ (dashed), and $t=2000$ (solid), at $T=0.5$. } \vspace{-0.6cm}
\end{centering}
\end{figure*}
\begin{figure*}
\begin{centering}
\includegraphics[width=14cm]{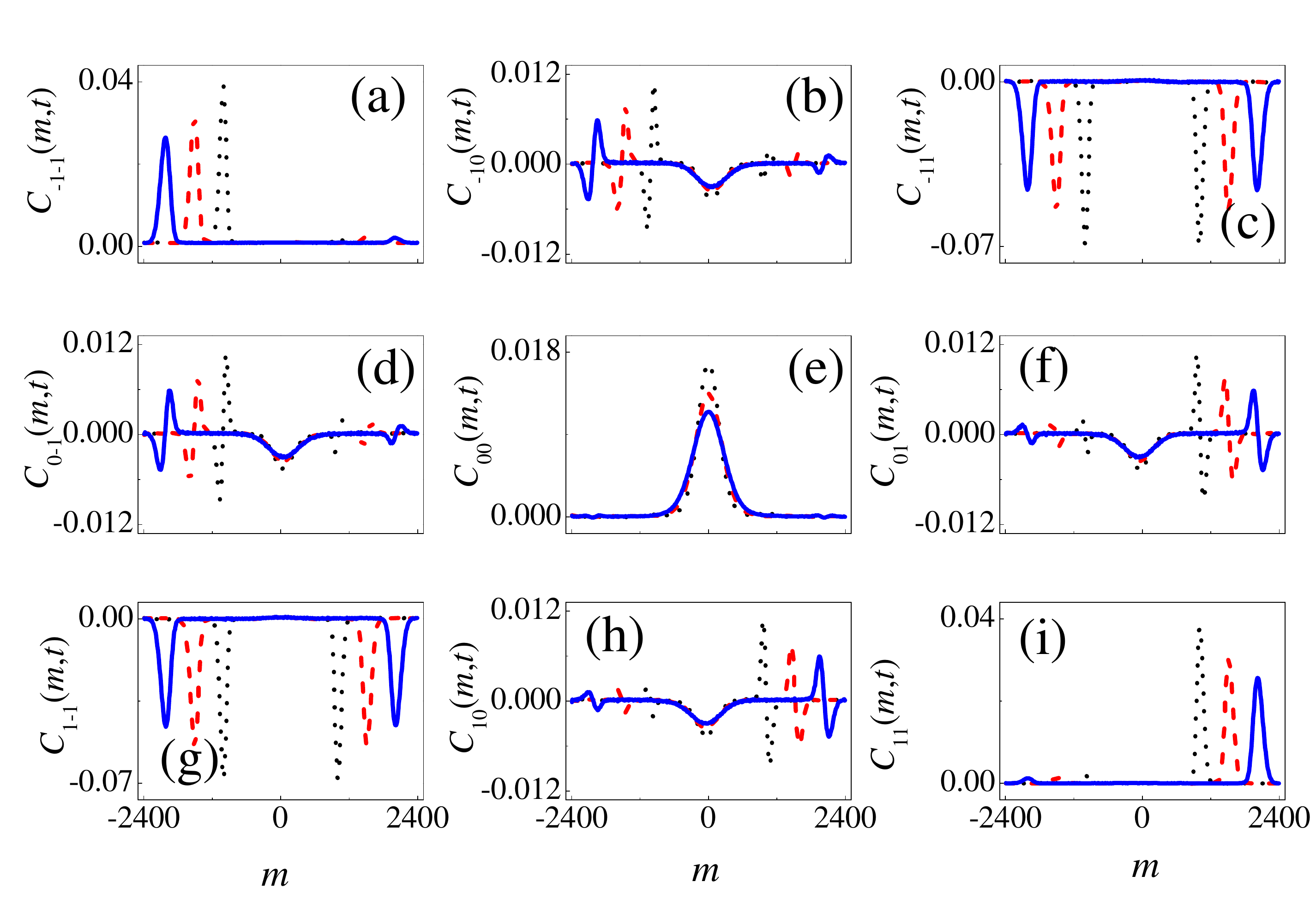} \vspace{-0.6cm}
\caption{\label{SFig6} The cubic-plus-quartic chain ($\alpha=1$): the $3 \times 3$ correlator matrix of hydrodynamic modes for three long times: $t=1000$ (dotted), $t=1500$ (dashed), and $t=2000$ (solid), at $T=0.5$. } \vspace{-0.6cm}
\end{centering}
\end{figure*}
\begin{figure*}
\begin{centering}
\includegraphics[width=14cm]{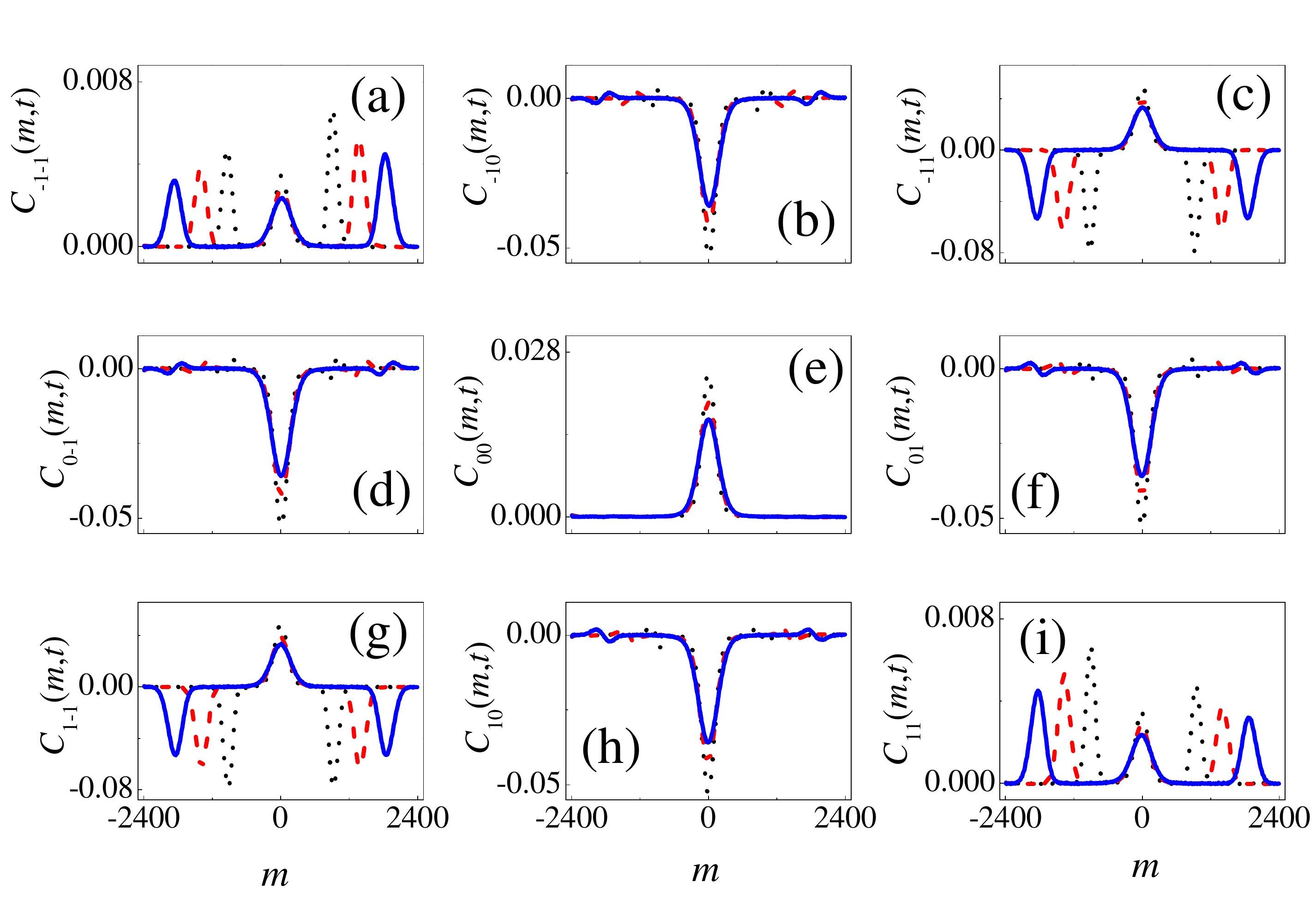} \vspace{-0.6cm}
\caption{\label{SFig7} The cubic-plus-quartic chain ($\alpha=2$): the $3 \times 3$ correlator matrix of hydrodynamic modes for three long times: $t=1000$ (dotted), $t=1500$ (dashed), and $t=2000$ (solid), at $T=0.5$. } \vspace{-0.6cm}
\end{centering}
\end{figure*}
\begin{figure*}
\begin{centering}
\includegraphics[width=14cm]{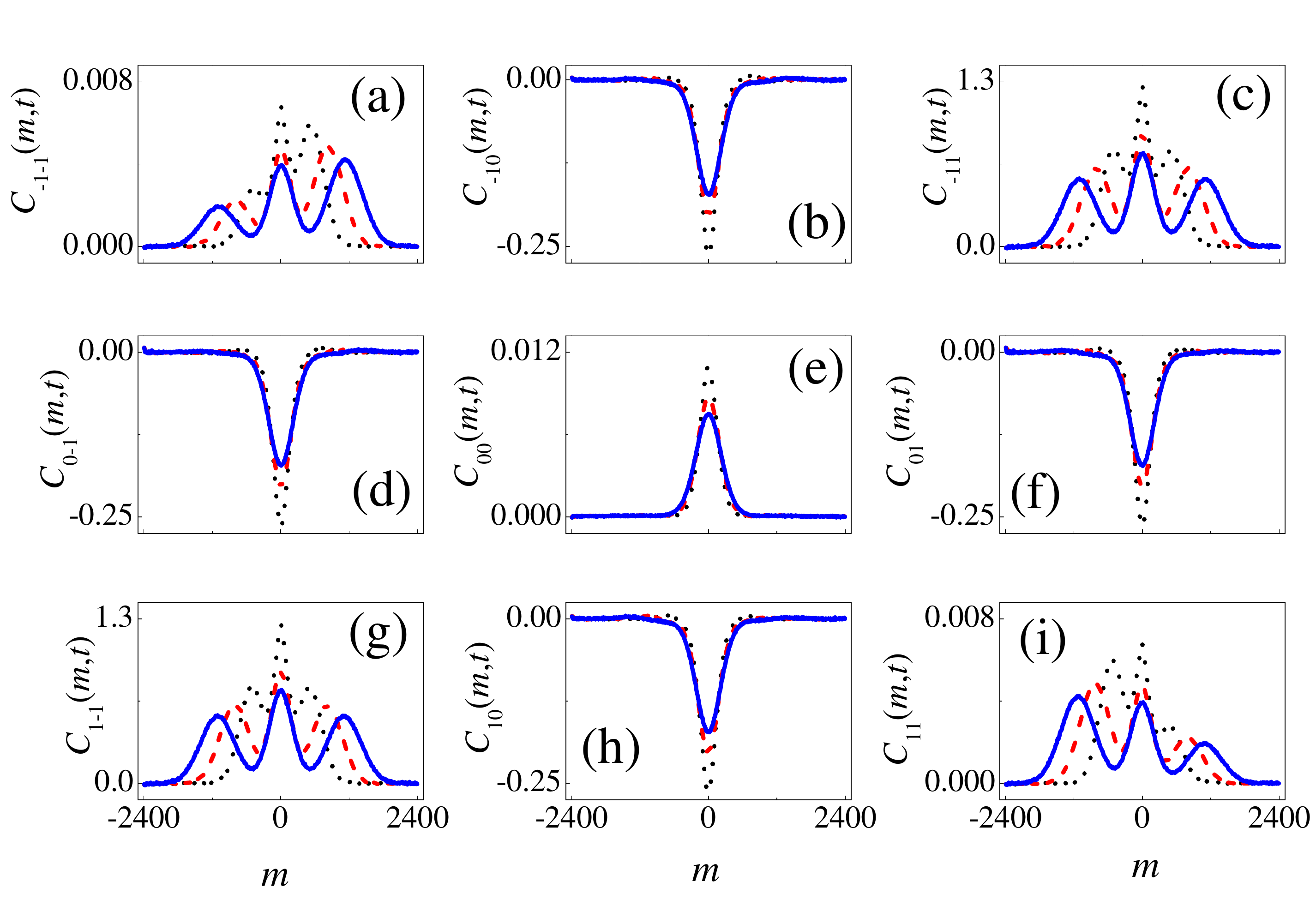} \vspace{-0.6cm}
\caption{\label{SFig8} The cubic-plus-quartic chain ($\alpha=3$): the $3 \times 3$ correlator matrix of hydrodynamic modes for three long times: $t=1000$ (dotted), $t=1500$ (dashed), and $t=2000$ (solid), at $T=0.5$. } \vspace{-0.6cm}
\end{centering}
\end{figure*}
\section{IV. The cubic-plus-quartic chains}
To see the the mirror symmetry of hydrodynamic modes, in Figs.~\ref{SFig5},~\ref{SFig6},~\ref{SFig7}, and~\ref{SFig8} we also provide some typical results of the $3 \times 3$ correlator matrices of hydrodynamic modes for the cubic-plus-quartic chains.
\begin{figure*}
\begin{centering}
\includegraphics[width=12cm]{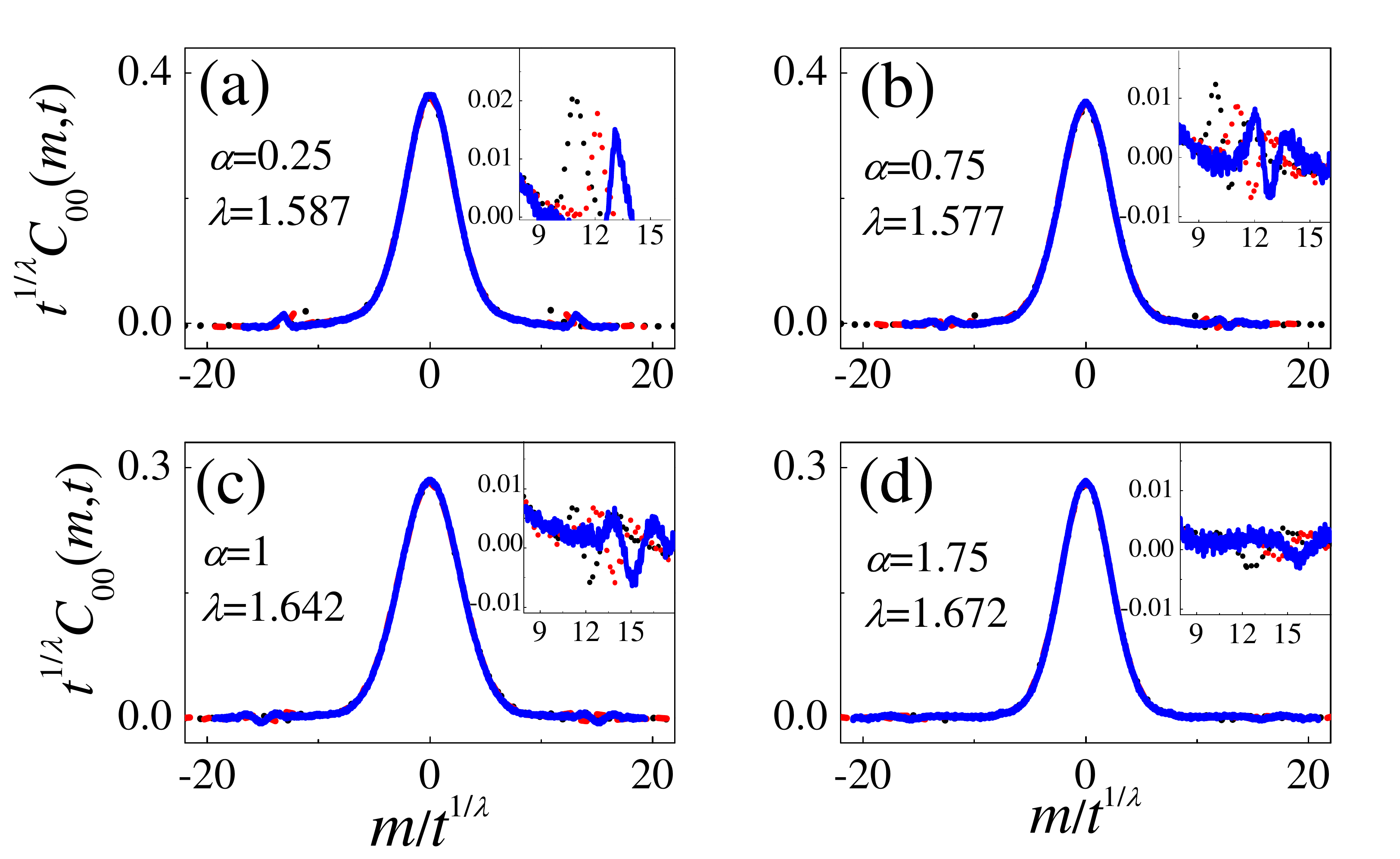} \vspace{-0.6cm}
\caption{\label{SFig9} Rescaled $C_{00}(m,t)$ ($T=0.5$) for the cubic-plus-quartic chains (see also [16]): (a) $\alpha=0.25$, (b) $\alpha=0.75$, (c) $\alpha=1$, and (d) $\alpha=1.75$. The insets show a zoom of the side peaks (right).} \vspace{-0.6cm}
\end{centering}
\end{figure*}
\begin{figure*}
\begin{centering}
\includegraphics[width=12cm]{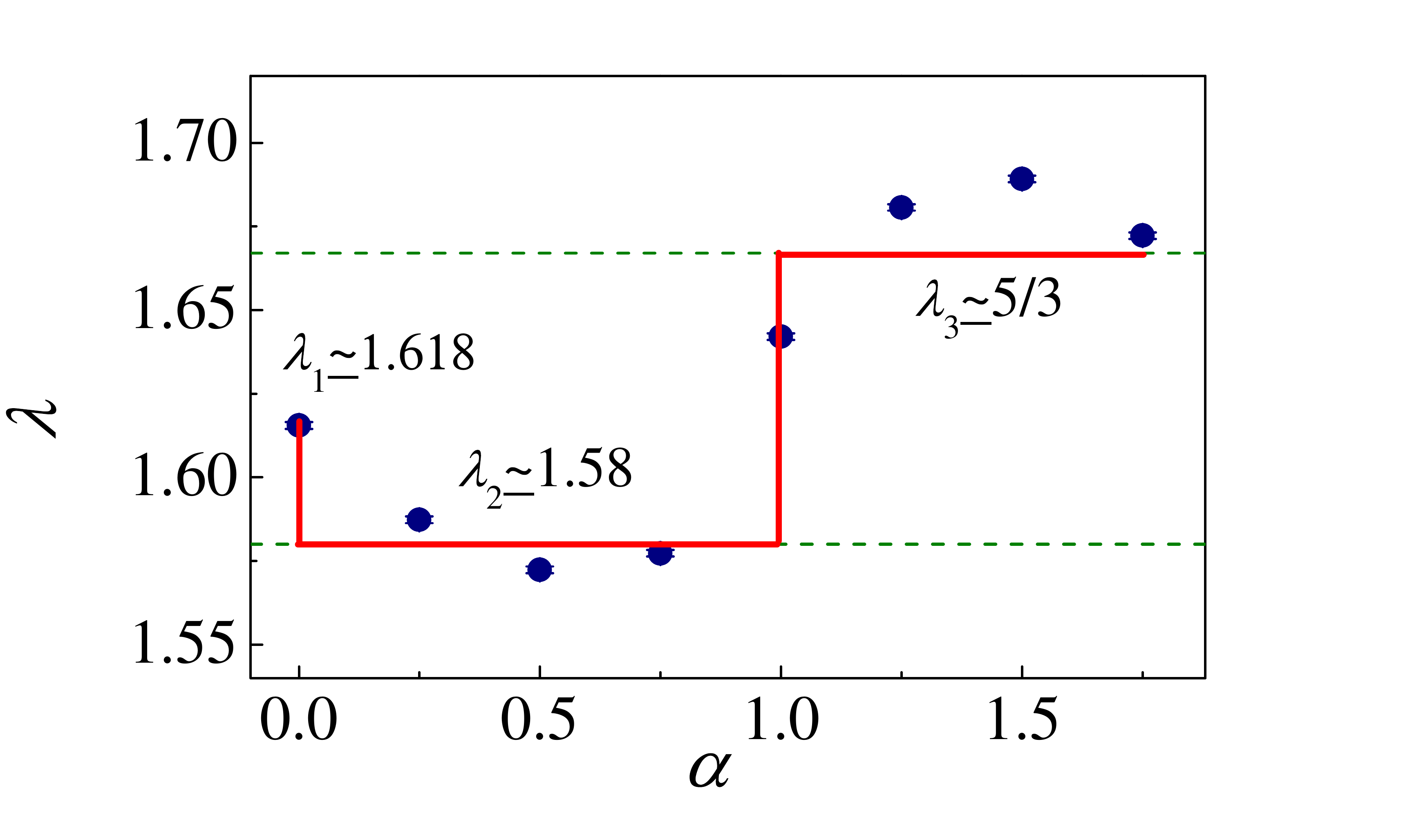} \vspace{-0.6cm}
\caption{\label{SFig10} $\lambda$ vs $\alpha$ for the cubic-plus-quartic chains roughly indicating three universality classes (see also [16]).} \vspace{-0.6cm}
\end{centering}
\end{figure*}
\begin{figure*}
\begin{centering}
\includegraphics[width=12cm]{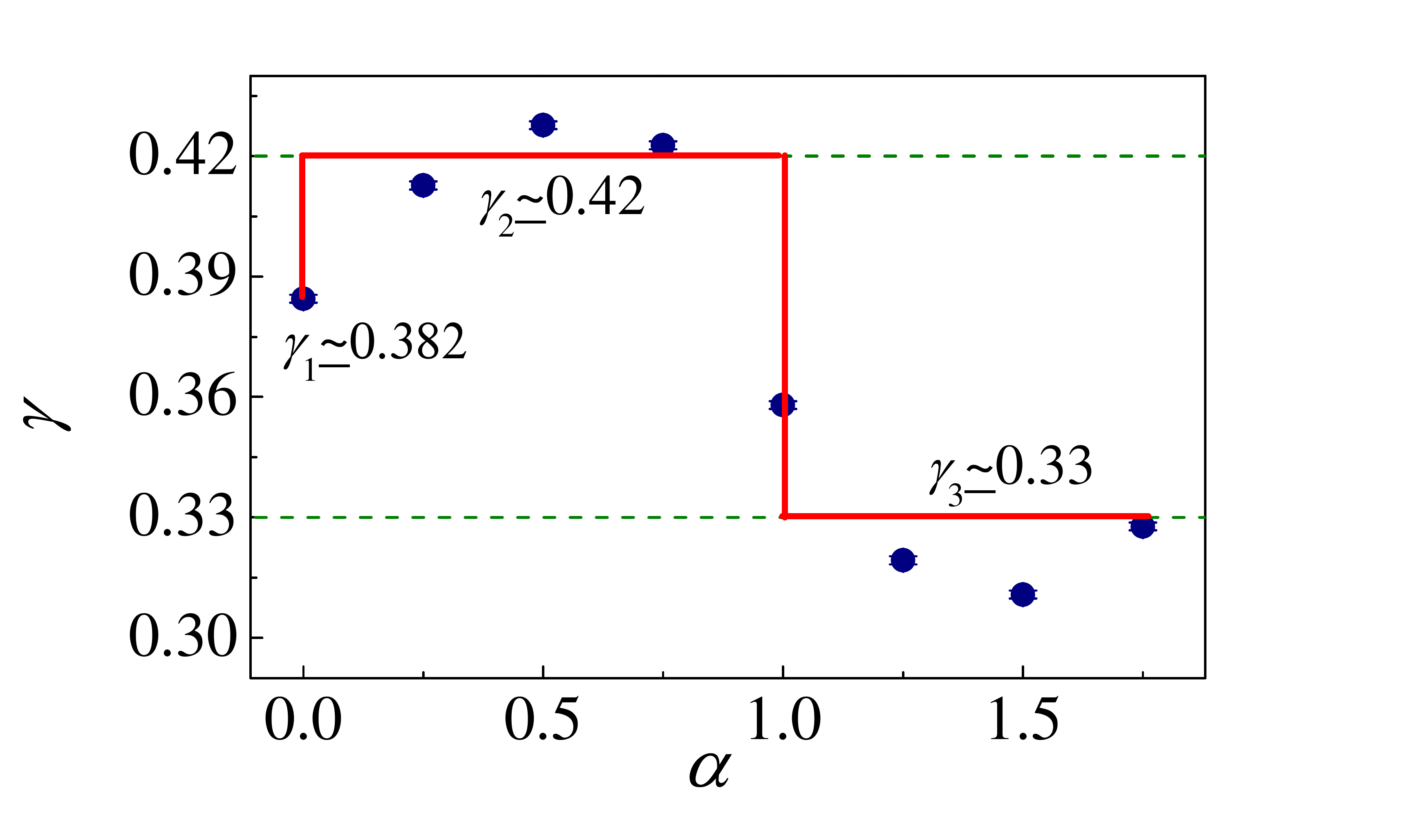} \vspace{-0.6cm}
\caption{\label{SFig11} $\gamma$ vs $\alpha$ plotted from the data of Fig.~\ref{SFig10} combined with Eq.~\eqref{Srelation}.} \vspace{-0.6cm}
\end{centering}
\end{figure*}
\section{V. Superdiffusive heat mode and three universality classes}
The nonlinearity-dependent relaxation of heat mode in the cubic-plus-quartic chains has already been explored in our previous work~[16]. Usually, this can be studied by using the following scaling formula
\begin{equation} \label{SScaling}
t^{1/\lambda} C_{00}(m,t) \simeq C_{00}(t^{-1/\lambda}m,t),
\end{equation}
with which a scaling exponent $\lambda$ can be obtained and related to the system size dependent exponent $\gamma$ of the thermal conductivity by
\begin{equation} \label{Srelation}
\gamma=2-\lambda.
\end{equation}
Figure~\ref{SFig9} depicts the rescaled $C_{00}(m,t)$ according to Eq.~\eqref{SScaling} and Fig.~\ref{SFig10} shows the $\alpha$-dependent $\lambda$. As seen, at least three universality classes of $\lambda$ can be roughly identified. This is consistent with the observed three classes of hydrodynamic modes couplings shown in Fig.~3: (i) $\alpha=0$, the couplings are only in the location of sound modes and there appears a universal time decaying law $t^{-1}$ [see Fig.~3(a)]; (ii) $0<\alpha<1$, the main couplings are still in the location of sound modes but they are distorted and long-time tails can be discovered [see Fig.~3(b)]; (ii) $\alpha \geq 1$, the couplings in the location of heat mode emerge [see Figs.~3(c-e)] and will be finally concentrated there [see Figs.~3(f)]. Finally, according to Eq.~\eqref{Srelation} one might infer at least three universal classes of $\gamma$ [see Fig.~\ref{SFig11}].
\section{VI. Results of the previous scheme}
\begin{figure*}
\begin{centering}
\includegraphics[width=14cm]{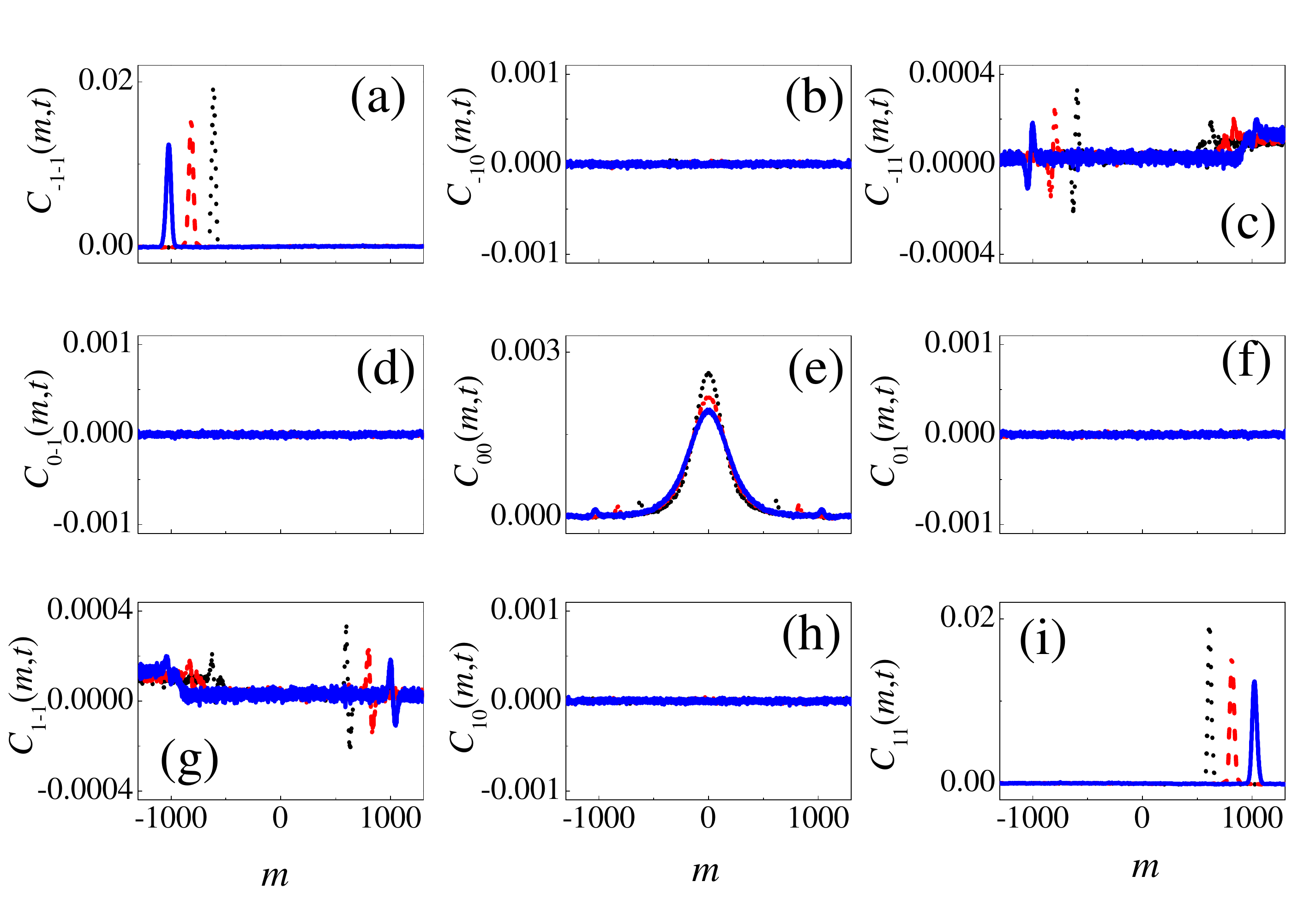} \vspace{-0.6cm}
\caption{\label{SFig12} The cubic-plus-quartic chain ($\alpha=0$, $T=0.5$): the $3 \times 3$ correlator matrix of hydrodynamic modes for three long times: $t=600$ (dotted), $t=800$ (dashed), and $t=1000$ (solid), derived from the previous scheme~[7, 11]~(see also [16]). } \vspace{-0.6cm}
\end{centering}
\end{figure*}

\begin{figure*}
\begin{centering}
\includegraphics[width=14cm]{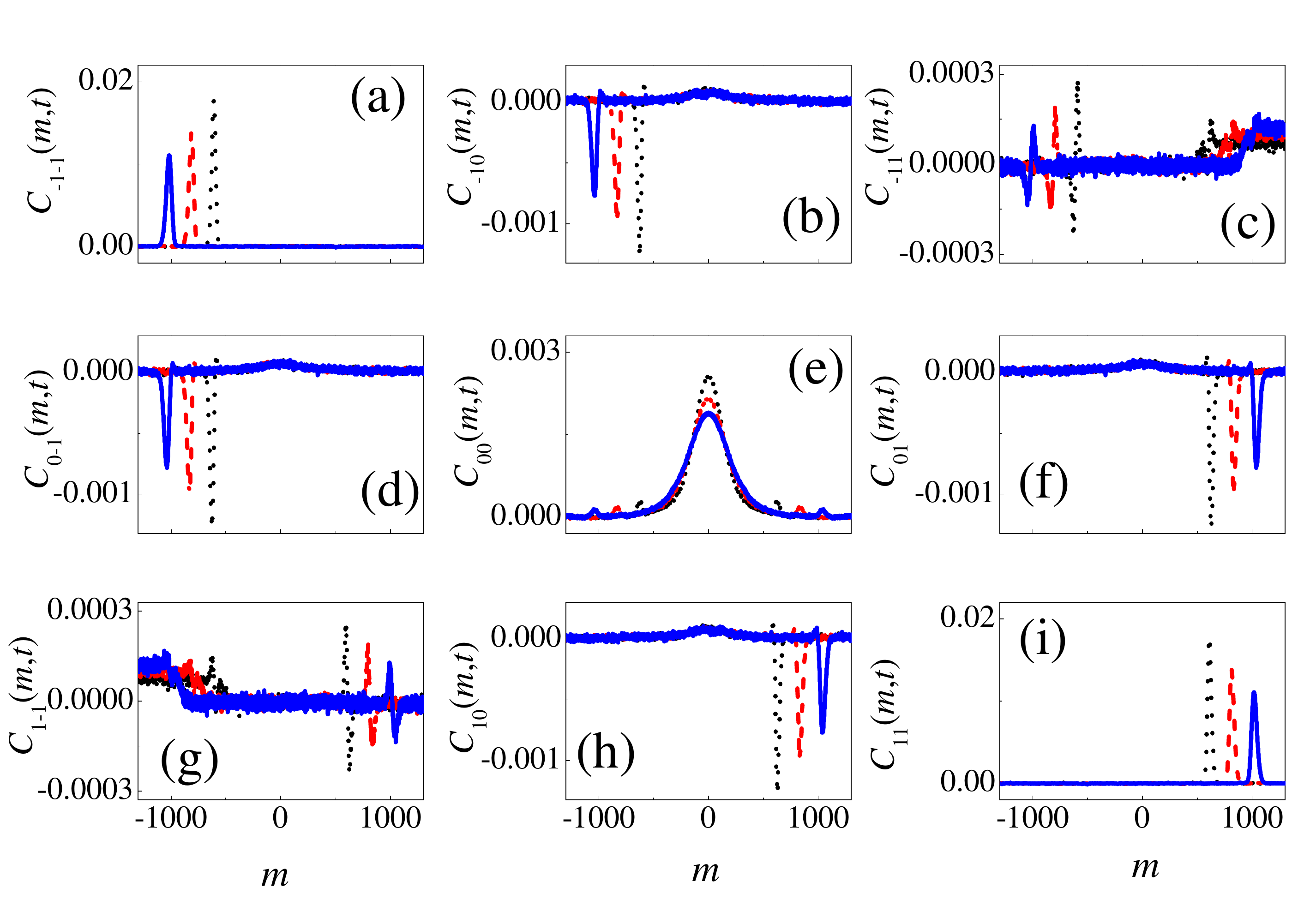} \vspace{-0.6cm}
\caption{\label{SFig13} The cubic-plus-quartic chain ($\alpha=0.25$, $T=0.5$): the $3 \times 3$ correlator matrix of hydrodynamic modes for three long times: $t=600$ (dotted), $t=800$ (dashed), and $t=1000$ (solid), derived from the previous scheme~[7, 11]~(see also [16]). } \vspace{-0.6cm}
\end{centering}
\end{figure*}

Finally, for the reader's reference we also provide two examples of the $3 \times 3$ correlator matrix of hydrodynamic modes derived from the previous scheme~[7, 11]. Figure~\ref{SFig12} depicts the results for the cubic-plus-quartic chain with $\alpha=0$ and Fig.~\ref{SFig13} plots the results for $\alpha=0.25$. As seen, the results are different from those obtained from our scheme. In particular, no cross-correlations are detected for $\alpha=0$; strange negative cross-correlations are observed for $\alpha \neq 0$. The reason for this difference might be that, to detect the hydrodynamic modes, the previous scheme always relies on a normal transformation~[11], which leads the estimation not direct. However, our scheme is directly from molecular dynamics.


\begin{thebibliography}{100}

\bibitem{FHD} J. O. de Zarate and J. Sengers, Hydrodynamic Fluctuations in Fluids and Fluid Mixtures (Elsevier, New York, 2006).

\bibitem{Noteadd} These noise terms are based on the flucutation-dissipation relation and can be obtained by a semi-microscopic kinetic theory, see for example ``J. R. Dorfman and E. G. D. Cohen, Difficulties in the kinetic theroy of Dense Gases, J. Math. Phys. \textbf{8}, 282 (1967).''

\bibitem{Longtail} B. J. Alder and T. E. Wainwright, Velocity Autocorrelations for Hard Spheres, Phys. Rev. Lett. \textbf{18}, 988 (1967).

\bibitem{FHD-1} M. H. Ernst, E. H. Hauge, and J. M. J. van Leeuwen, Asymptotic time behavior of correlation functions. II. Kinetic and potential terms, J. Stat. Phys. \textbf{15}, 7 (1976).

\bibitem{Narayan2002}
O. Narayan and S. Ramaswamy, Anomalous Heat Conduction in One-Dimensional Momentum-Conserving Systems, Phys. Rev. Lett.~\textbf{89}, 200601
(2002).

\bibitem{Beijeren2012} H. van Beijeren, Exact Results for Anomalous Transport in One-Dimensional Hamiltonian Systems, Phys. Rev. Lett.~\textbf{108},
180601 (2012).

\bibitem{Spohn2014} H. Spohn, Nonlinear fluctuating hydrodynamics for anharmonic
chains, J. Stat. Phys.~\textbf{154}, 1191 (2014).

\bibitem{Book1}
S. Lepri, \emph{Thermal Transport in Low Dimensions: From
Statistical Physics to Nanoscale Heat Transfer} (Springer, 2016).

\bibitem{LepriReport} S. Lepri, R. Livi, and A. Politi, Thermal conduction in classical
low-dimensional lattices, Phys. Rep.~\textbf{377}, 1 (2003).

\bibitem{DharReport}
A. Dhar, Heat transport in low-dimensional systems, Adv. Phys.~\textbf{57}, 457 (2008).






\bibitem{NumTest} S. G. Das, A. Dhar, K. Saito, C. B. Mendl, and H. Spohn, Numerical test of hydrodynamic fluctuation theory in the Fermi-Pasta-Ulam chain, Phys. Rev. E.~\textbf{90}, 012124 (2014).

\bibitem{NFHD-no-3} D. S. Sato, Pressure-induced recovery of Fourier's law in one-dimensional momentum-conserving systems, Phys. Rev. E.~\textbf{94}, 012115 (2016).

\bibitem{Generalized-1} B. Doyon, T. Yoshimura, and J. Caux, Soliton Gases and Generalized Hydrodynamics, Phys. Rev. Lett. \textbf{120}, 045301 (2018).

\bibitem{Generalized-2} M. Fagotti, Higher-order generalized hydrodynamics in one dimension: The noninteracting test, Phys. Rev. B \textbf{96}, 220302(R) (2017).

\bibitem{Livi2020} R. Livi, Heat transport in one dimension, J. Stat. Mech. (2020) 034001.

\bibitem{Xiong2018} D. Xiong, Observing golden-mean universality class in the scaling of thermal transport, Phys. Rev. E \textbf{97}, 022116 (2018).

\bibitem{Lee-Dadswell2008} G. R. Lee-Dadswell, B. G. Nickel, and C. G. Gray, Detailed Examination of Transport Coefficients in Cubic-Plus-Quartic Oscillator Chains,  J. Stat. Phys. \textbf{132}, 1 (2008).

\bibitem{SM} See Supplemental Material for the detailed $3 \times 3$ correlator matrices of hydrodynamic modes in the FPU-$\beta$, FPU-$\alpha$$\beta$, and the cubic-plus-quartic chains, and relevant scaling information of heat mode, obtained from our scheme. For the reader's refernece, we also provide two typical correlator matrices derived from the previous scheme~\cite{Spohn2014,NumTest} for comparison.

\bibitem{Chen2013} S. Chen, Y. Zhang, J. Wang, and H. Zhao, Diffusion of heat, energy, momentum, andmass in one-dimensional systems, Phys. Rev. E \textbf{87}, 032153 (2013).


\bibitem{Nianbei2010} N. Li, B. Li, and S. Flach, Energy Carriers in the Fermi-Pasta-Ulam $\beta$ Lattice: Solitons or Phonons? Phys. Rev. Lett. \textbf{105}, 054102 (2010).


\bibitem{Xiong2016} This has been verified for the system of interest, while we also note that for some special systems, it may not always hold, see for example ``D. Xiong, Underlying mechanisms for normal heat transport in one-dimensional anharmonic oscillator systems with a double-well interparticle interaction, J. Stat. Mech. (2016) 043208.''

\bibitem{Notenew} The couplings decay as $t^{-1}$, which is just one-order faster than the decay $t^{-\frac{1}{2}}$ of sound modes. In this sense, the decoupling hypothesis claimed by NFHD is almost valid.

\bibitem{Rotor-1} C. Giardin\`{a}, R. Livi, A. Politi, and M. Vassalli, Finite Thermal Conductivity in 1D Lattices, Phys. Rev. Lett. \textbf{84}, 2144 (2000).


\bibitem{Rotor-2} O. V. Gendelman and A. V. Savin, Normal Heat Conductivity of the One-Dimensional Lattice with Periodic Potential of Nearest-Neighbor Interaction, Phys. Rev. Lett. \textbf{84}, 2381 (2000).

\bibitem{Rotor-3} H. Spohn, Fluctuating hydrodynamics for a chain of nonlinearly coupled rotators, arXiv:1411.3907.

\bibitem{Rotor-4} S. G. Das and A. Dhar, Role of conserved quantities in normal
heat transport in one dimension, arXiv:1411.5247.

\bibitem{Xiong2020} S. You, D. Xiong, and J. Wang, Thermal rectification in the thermodynamic
limit, Phys. Rev. E \textbf{101}, 012125 (2020). Here the role of
pressure fluctuations are used to achieve a highly efficient thermal
rectification in the thermodynamic limit.


\bibitem{NFH3D} R. Barreto, M. F. Carusela, and A. G. Monastra, Nonlinear fluctuating hydrodynamics with many conserved fields: The case of a three-dimensional anharmonic chain, Phys. Rev. E 100, 022118 (2019).



\end{thebibliography}
\end{document}